\documentclass[preprints,article,accept,moreauthors,pdftex,a4paper]{Definitions/mdpi}

\firstpage{1} 
\pubvolume{1}
\issuenum{1}
\articlenumber{0}
\pubyear{2021}
\copyrightyear{2020}
\datereceived{} 
\dateaccepted{} 
\datepublished{} 
\hreflink{https://doi.org/} %
\pdfoutput=1

\usepackage{csquotes}
\usetikzlibrary{positioning,arrows,shapes,decorations.pathmorphing,spy,external}
\usepackage{pgfplots}
\usepgfplotslibrary{groupplots}
\usepackage{mathtools}
\usepackage[output-decimal-marker={.},
  number-unit-product={\,},group-separator={\,},
  group-minimum-digits=4,range-phrase=--]{siunitx}
\usepackage{varwidth}
\usepackage{basicmath}
\usepackage{eqparbox}
\usepackage[noend]{algpseudocode}

\DeclareFloatingEnvironment[name=Algorithm]{algoenv}
\algdef{SE}[LOOP]{Loop}{EndLoop}[1]{\algorithmicloop\ #1:}{\algorithmicend\ \algorithmicloop}
\algtext*{EndLoop}

\newcommand*\patchAmsMathEnvironmentForLineno[1]{%
  \expandafter\let\csname old#1\expandafter\endcsname\csname #1\endcsname
  \expandafter\let\csname oldend#1\expandafter\endcsname\csname end#1\endcsname
  \renewenvironment{#1}%
     {\linenomath\csname old#1\endcsname}%
     {\csname oldend#1\endcsname\endlinenomath}}%
\newcommand*\patchBothAmsMathEnvironmentsForLineno[1]{%
  \patchAmsMathEnvironmentForLineno{#1}%
  \patchAmsMathEnvironmentForLineno{#1*}}%
\AtBeginDocument{%
\patchBothAmsMathEnvironmentsForLineno{equation}%
\patchBothAmsMathEnvironmentsForLineno{align}%
\patchBothAmsMathEnvironmentsForLineno{flalign}%
\patchBothAmsMathEnvironmentsForLineno{alignat}%
\patchBothAmsMathEnvironmentsForLineno{gather}%
\patchBothAmsMathEnvironmentsForLineno{multline}%
}

\pgfplotsset{compat=1.14}

\Title{Blind Source Separation in Polyphonic Music Recordings Using Deep Neural Networks Trained via Policy Gradients}

\TitleCitation{Blind Source Separation in Polyphonic Music Recordings Using Deep Neural Networks Trained via Policy Gradients}

\Author{Sören Schulze$^{1,}$*\orcidA{}, Johannes Leuschner$^{1}$\orcidB{} and Emily J. King$^{2}$\orcidC{}}

\AuthorNames{Sören Schulze, Johannes Leuschner and Emily J. King}

\AuthorCitation{Schulze, S.; Leuschner, J.; King, E.J.}
\address{%
$^{1}$ \quad Center for Industrial Mathematics, University of Bremen, Bibliothekstr. 5, 28359 Bremen, Germany; sschulze@uni-bremen.de (S.S.); jleuschn@uni-bremen.de (J.L.)\\
$^{2}$ \quad Mathematics Department, Colorado State University, 1874 Campus Delivery, 111 Weber Bldg, 80523 Fort Collins, CO, USA; emily.king@colostate.edu (E.J.K.)}

\corres{sschulze@uni-bremen.de}

\abstract{
  We propose a method for the blind separation of sounds of musical instruments
  in audio signals.  We describe the individual tones via a parametric model,
  training a dictionary to capture the relative amplitudes of the harmonics.
  The model parameters are predicted via a U-Net, which is a type of
  deep neural network.  The network is trained without ground truth
  information, based on the difference between the model prediction and the
  individual time frames of the short-time Fourier transform.
  Since some of the model parameters do not
  yield a useful backpropagation gradient, we model them stochastically and
  employ the policy gradient instead.  To provide phase information and
  account for inaccuracies in the dictionary-based representation, we
  also let the network output a direct prediction, which we then use
  to resynthesize the audio signals for the individual instruments.
  Due to the flexibility of the neural network, inharmonicity can be
  incorporated seamlessly and no preprocessing of the input spectra
  is required.  Our algorithm yields high-quality separation results with
  particularly low interference on a variety of different audio samples,
  both acoustic and synthetic, provided that the sample contains enough
  data for the training and that the spectral
  characteristics of the musical instruments are sufficiently stable
  to be approximated by the dictionary.}


\begin{document}

\section{Introduction}

We address the problem of unmixing the contributions of multiple different
musical
instruments from a single-channel audio recording.  We assume that each
instrument only plays a single musical tone at a time and that the sound
of the instruments follows a stationary tone model aimed at
woodwind, brass, and string instruments.

Since we perform \emph{blind} separation, we do not make any
prior assumptions specific to the sounds of the individual instruments,
but we distinguish them based on the proximity to the entries of a
\emph{dictionary} which we learn in the process.

For the time-frequency representation of the audio signals, we use
the sampled complex-valued output of the short-time Fourier transform,
which can be interpreted as the analysis coefficients of a Gabor frame.
This representation has the advantage of being perfectly linear and
easy to project back to a time-domain signal, but it is not
\emph{pitch-invariant}; that is, the distance of the frequency axis
corresponding to a certain musical interval varies based on the pitch
of the tones.

The problem of identifying the pitch of the tones is non-convex on a
global scale and possesses a large number of local minima.
Therefore, general numerical optimization methods are
not appropriate.  Instead, we predict the parameters via a
\emph{U-Net} \cite{Ronneberger2015}, which is a type of deep neural network.
For the problematic parameters like pitch, we use policy gradients for training,
which is a technique originating from deep reinforcement learning
\cite[cf.][]{Sutton20}.

\subsection{Related Work}

The audio source separation problem can be formulated in a variety
of settings.  (See \cite{Vincent18,Makino18,Chien18} for a thorough
overview.)  For the purpose of this work, we only regard the
case that the input signal is single-channel and the separation is
therefore always underdetermined.  Different algorithms can be used if
multiple channels are available (typically corresponding to microphones
simultaneously recording the audio scene).  Also, we always assume
melodic instruments rather than speech or percussive instruments.
While different kinds of prior
information can be considered (such as specific characteristics of the sounds
of the instruments, training data, or the musical score), we concentrate
on the blind case with no prior information and instead rely on a learned
parametric model for the sounds of the individual instruments.

Many algorithms for this problem are based on
non-negative matrix factorization (NMF) of the spectrogram.
In the simplest form, each tone
of an instrument at a particular pitch has its own representation
as a dictionary atom \cite{Smaragdis03,Plumbley05}.
To make the representation of the sound of a particular
instrument applicable at arbitrary pitch (\emph{pitch-invariance}),
one often employs \emph{tensor factorization} \cite[cf.][]{Chien18}.
In this case, the use of a log-frequency
spectrogram (such as the constant-Q transform or the mel spectrogram)
can be helpful, since it is also pitch-invariant in the sense that changing
the fundamental frequency of a tone merely causes a shift in the representation
\cite{Fitzgerald05a,Jaiswal11a,Jaiswal11b,Jaiswal13}.

A separation approach that is mathematically equivalent to NMF is
\emph{probabilistic latent component analysis} (PLCA)
\cite{Smaragdis06,Smaragdis07} which also exists in variants that use a pitch-invariant model
on top of a log-frequency spectrogram \cite{Smaragdis08,Fuentes11}.
The next step in abstraction is to model the individual harmonics separately
while enforcing sparsity in the spectral representation \cite{Fuentes13}.

Decreasing the variability, e.g., the number of parameters,
in a model is generally beneficial since it reduces
the risk of overfitting.  However, we can go one step further by employing
an explicit physical model for the tones of the instruments involving a
minimum number of parameters.  This has the additional advantage
that while the previously discussed approaches with a log-frequency
spectrogram can only be pitch-invariant if the instruments are tuned to
the same log-frequency scale (\emph{equal-temperament tuning}
\cite[cf.][]{Neuwirth97}\footnote{%
  Equal temperament is defined as the frequency ratio corresponding
  to a musical interval being constant regardless of pitch.}),
a physical model can be evaluated at any fundamental frequencies
and sampled arbitrarily.  This is crucial when dealing with acoustic
instruments that either deliberately deviate from equal temperament
or might simply be slightly out of tune.

With a continuous model, it is thus possible to use a pitch-invariant
representation on a
linear-frequency spectrogram.  However, the challenge then is to identify
the fundamental frequencies on a continuous domain.  Duan et al.\ \cite{Duan08}
use a peak detection and clustering algorithm to reduce the problem to a
combinatorial one that can be approached via appropriate heuristics.
Hennequin et al.\ \cite{Hennequin10}, in a polyphonic single-instrument setting,
consider the fundamental frequency as an optimizable parameter and use
an NMF-type update rule, but they remark that this is only possible on a
local scale due to the high number of local minima.

Even with a physical model, it turns out that the
log-frequency spectrogram can still be helpful:
Schulze and King \cite{Schulze21} use its pitch-invariance property to
obtain the approximate fundamental frequency of a tone via simple
cross-correlation.  After that, numerical optimization is performed
to improve the estimate on a local scale.
The optimization procedure also incorporates inharmonicity, which violates
pitch-invariance, and variable width of the peaks in the spectrum
which can occur, for instance, at tone boundaries.
However, as explained there, using a log-frequency spectrogram inevitably
results in a loss in frequency resolution which in this case needs to be
mitigated via heavy preprocessing.  Also, phase information is lost completely.
Therefore, if given the choice, we argue that a linear-frequency
representation should be preferred.

Due to the general success of deep neural networks, it is not surprising
that they have also been applied to audio source separation problems.  In
fact, when it comes to supervised separation (with labeled training data),
they dominate the state of the art
\cite{Stoeter19,Defossez19,Li21,Nachmani20,Takahashi21}.
While supervised training is the
\enquote{classical} way in which neural networks are used, it was demonstrated
by Ulyanov el al.\ with the \emph{deep image prior} (DIP) approach
\cite{Ulyanov18} that the structure of (convolutional) neural networks is
inherently useful for representing natural images.  This technique was used
for image decomposition by Gandelsman et al.\ via the \emph{double-DIP}
algorithm \cite{Gandelsman19}.  Given this success, it was natural to also
apply this method to audio data, leading to the \emph{deep audio prior}
approach by Tian et al.\ \cite{Tian19}.  Based on this, Narayanaswamy et al.\
\cite{Narayanaswamy20} used \emph{generative adversarial networks} (GANs)
trained on unlabeled training data as priors, further improving the
quality of the output signals.

The problem with all the previously presented separation algorithms
based on unsupervised training of neural networks is that they make the
assumption that the separated signals are stochastically independent.
This case is often referred to as the \emph{cocktail party problem}, but
it is different from polyphonic music, in which the tones are usually
both rhythmically and harmonically aligned.  Therefore, rather than relying
on general stastistical properties of the signals, we use deep neural networks
in conjunction with a parametric model.

\emph{Policy gradients} were pioneered within reinforcement learning
with the \emph{REINFORCE} algorithm \cite{Williams92}, which is designed
to train a neural network to predict a discrete variable.  While reinforcement
learning has progressed towards \emph{actor-critic} methods
\cite[cf.][]{Sutton20} and, famously, the \emph{AlphaGo Zero}
\cite{Silver17}, \emph{AlphaZero} \cite{Silver18},
and \emph{MuZero} \cite{Schrittwieser20} algorithms
based on \emph{Monte Carlo tree search} (MCTS),
we stay relatively close to the original approach, but we extend the
formulation by adding deterministic values, combining policy gradients
with backpropagation gradients.

\section{Data Model}

\subsection{Tone Model}

An idealized model for the tones of
woodwind, brass, and string instruments is that of the wave
equation, which is a hyperbolic 2nd-order partial differential equation.
However, for certain string instruments, the stiffness
in the strings is non-negligible,
and this leads to the introduction of 4th-order terms which cause
inharmonicity \cite[cf.][]{Fletcher98}.
The corresponding solution consists of real-valued
sinusoids, which consequently also appear in the audio signal.
However, due to
\begin{equation}
  \sin(\xi)=\frac{e^{i\xi}-e^{-i\xi}}{2i},\qquad
  \cos(\xi)=\frac{e^{i\xi}+e^{-i\xi}}{2},
\end{equation}
we can also express them as complex exponentials.  Since we do not
need the negative exponential, our model of a tone of a musical instrument
is as follows:
\begin{equation}\label{eq:timemodel}
x(t)=\sum_{h}a_h\,e^{i2\pi f_ht},
\end{equation}
with
\begin{equation}
  f_h=f_1^\circ h\sqrt{1+bh^2},
\end{equation}
where $h=1,\dotsc,N_{\mathrm{har}}$ are the \emph{harmonics}
(with $N_{\mathrm{har}}\in\N$),
$a_h\in\C$ is the complex \emph{amplitude},
$f_h$ is the \emph{frequency} for the specific harmonic,
$f_1^\circ>0$ is the \emph{fundamental frequency} of the tone,
and $b\geq0$ is the \emph{inharmonicity}.

For illustration, an artificial application of the tone model is
provided in Figure~\ref{fig:tonemodel}.  While the inharmonicity is
exaggerated in comparison to real acoustic pianos, the increase in
distance between the harmonics in the frequency domain is clearly visible.
In the time domain, this has the effect that the overall signal is no longer
periodic, even though the signals stemming from the individual harmonics are.

\end{paracol}
\nointerlineskip
\begin{figure}
  \widefigure
  \small\centering
  \tikzsetnextfilename{Figure_\the\numexpr(\thefigure+1)\relax}
  \begin{tikzpicture}
    \begin{groupplot}[group style={group size=2 by 1,horizontal sep=0.075\textwidth},
        width=0.48\textwidth,title style={yshift=-1ex},
        xlabel={Time [$\si{ms}$]},
        height=5cm,enlargelimits=false,
        legend cell align=left]
      \nextgroupplot[xmin=0,xmax=20,ymin=1e-5,ymax=1,xlabel={Frequency [$\si{kHz}$]},title={Harmonic amplitudes}]
      \addplot[ycomb,thick,mark=*] table {tonesig-spect.dat};
      \nextgroupplot[ymin=-15,ymax=15,%
        legend pos=south east,
        title={Time-domain signal}]
      \addplot[line join=bevel] table[y expr=\thisrowno{1}] {tonesig-time.dat};
      \addlegendentry{Signal}
      \addplot[line join=bevel,red,thick] table[y expr=\thisrowno{2}] {tonesig-time.dat};
      \addlegendentry{Fundamental mode ($h=1$)}
    \end{groupplot}
  \end{tikzpicture}
  \caption{Illustrative example for the signal model with
    a fundamental frequency of $f_1^\circ=\SI{440}{Hz}$ and
    an inharmonicity parameter of $b=\num{e-2}$}
  \label{fig:tonemodel}
\end{figure}
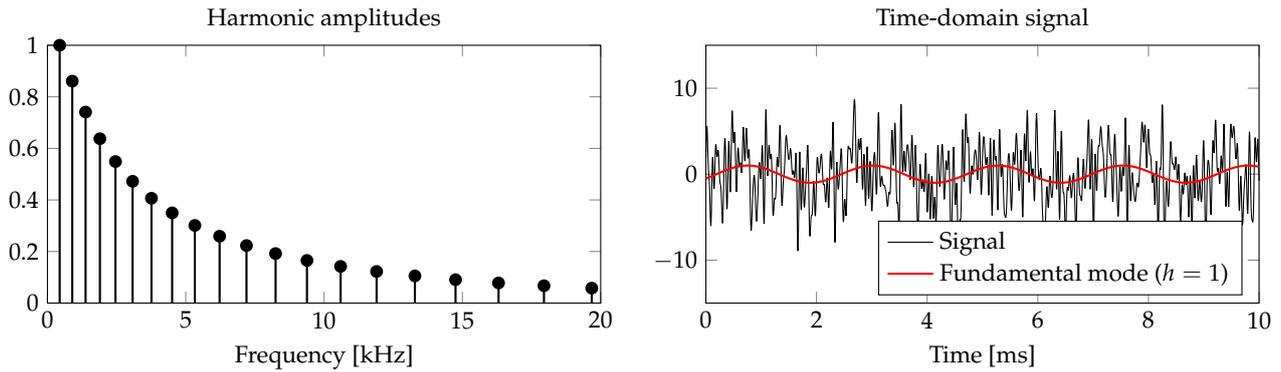
\begin{paracol}{2}
\makeatletter\ifthenelse{\equal{\@status}{submit}}{\linenumbers}{}\makeatother
\switchcolumn

\subsection{Time-Frequency Representation}

Due to the sinusoidal nature of the tone model \eqref{eq:timemodel},
it is advantageous to consider the signal in the frequency domain
rather than the time domain.  However, in reality, music is not
stationary over longer periods of time; tones start, end, and change
in volume or frequency.  Therefore, we only consider the frequency
spectrum of short excerpts in time, which is called a
\emph{time-frequency representation}.  We compute the
\emph{short-time Fourier transform} (STFT) of a signal $X\in L_2(\R)$
via:\footnote{In our notation, the uppercase letters $X,Z,Y$
always refer to measured data, while their lowercase counterparts
$x,z,y$ are our corresponding models.}
\begin{equation}
\mathcal{V}_wX(t,f)=\int_{-\infty}^\infty
X(\tau)\,w(\tau-t)\,e^{-i2\pi f\tau}\;\dif\tau,
\end{equation}
where $t,f\in\R$ correspond to the time and frequency axes and
$w\in L_2(\R)$ is the real-valued \emph{analysis window}
\cite[cf.][]{Groechenig01}.

We sample the representation as:
\begin{equation}\label{eq:sample}
  Z[k,l]\coloneqq \mathcal{V}_wX(\alpha k,\beta l)
  \,e^{i2\pi \alpha k \beta l},\qquad k,l\in\Z,
\end{equation}
with the time and frequency constants $\alpha,\beta>0$.
For $\alpha\beta<1$ and with the Gaussian window
\begin{equation}
  w(t)=\frac{1}{\sqrt{2\pi\zeta^2}}\exp\biggl(-\frac{t^2}{2\zeta^2}\biggr),\qquad
  \zeta>0,
\end{equation}
a so-called \emph{Gabor frame}
is formed, meaning that any function $X\in L_2(\R)$
is uniquely determined by its respective set of \emph{analysis coefficients}
$Z[k,l]$, $k,l\in\Z$ \cite[cf.][]{Groechenig01}.
In practice, we only consider a finite number of indices
$k=1,\dotsc,n_{\mathrm{len}}$ (for time) and
$l=0,\dotsc,n_{\mathrm{spc}}-1$ (for frequency)
with $n_{\mathrm{len}},n_{\mathrm{spc}}\in\N$, where we again neglect negative
frequencies.  If $X$ is real-valued, we have
$Z[k,-l]=\overline{Z[k,l]}$.

For the tone model \eqref{eq:timemodel}, the STFT yields:
\begin{equation}
  \mathcal{V}_wx(t,f)=
  \sum_{h} a_h\exp\Biggl(-\frac{(f-f_h)^2}{2\sigma^2}-i2\pi(f-f_h)t\Biggr),
\end{equation}
with $\zeta\sigma=1/(2\pi)$.\footnote{%
  For tones at very low frequencies, the negative frequencies that were
  omitted from the tone model \eqref{eq:timemodel} can cause interference
  with the positive part of the frequency spectrum.  With typical audio
  signals, this interference is not strong enough to become a problem.
  It would be straight-forward to add them back in, but
  this would increase the computational cost.
}
When sampling according to \eqref{eq:sample},
we obtain:
\begin{equation}\label{eq:tonespec}
  \begin{aligned}
  z[k,l]
  &\;=\mathcal{V}_wx(\alpha k,\beta l)\,e^{i2\pi \alpha k \beta l}\\
  &\;=\sum_{h} a_h\exp\Biggl(-\frac{(\beta l-f_h)^2}{2\sigma^2}-i2\pi(\beta l-f_h)\alpha k\Biggr)\,e^{i2\pi \alpha k \beta l}\\
  &=\sum_{h} a_h\exp\Biggl(-\frac{(\beta l-f_h)^2}{2\sigma^2}+i2\pi f_h\alpha k\Biggr).
  \end{aligned}
\end{equation}
Thus, for each harmonic $h$, the phase is constant for a fixed time
index $k$.

Our default choice, assuming a sampling frequency of
$f_\mathrm{s}=\SI{48}{kHz}$, is:
\begin{equation}\label{eq:default}
\zeta=\frac{1024}{f_\mathrm{s}}=\SI[parse-numbers=false]{21.\overline{3}}{ms},\quad
\alpha=\frac{\zeta}{2}=\frac{512}{f_\mathrm{s}}=\SI[parse-numbers=false]{10.\overline{6}}{ms},\quad
\beta=\frac{1}{12\zeta}=\frac{f_\mathrm{s}}{\num{12288}}=\SI{3.90625}{Hz}.
\end{equation}
The value of $\zeta$ is short enough to capture
rhythm, while the quantity $1/(2\pi\zeta)\approx\SI{7.46}{Hz}$, that will
become important later, is well below the fundamental frequencies considered.
Our choice of $\alpha$ ensures that Gaussian windows spaced by $\alpha$
overlap narrowly.

For practical computations, it is common to limit the support of
the window $w$ to $[-1/(2\beta),1/(2\beta)]$.  Due to $1/(2\beta)=6\zeta$,
our value for $\beta$ makes the resulting error negligible.

\subsection{Dictionary Representation}

In order to differentiate between instruments in a music recording,
we make the simplifying assumption that the tones for each instrument
$\eta=1,\dotsc,N_{\mathrm{ins}}$ (where $N_{\mathrm{ins}}$ is the total number
of instruments) follow
a characteristic pattern, namely that we can express the amplitudes of
the harmonics as:
\begin{equation}\label{eq:dictmodel}
  a_h=a\,D[h,\eta]\,e^{i\tilde{\varphi}_h},\qquad h=1,\dotsc,N_{\mathrm{har}},
\end{equation}
where $a\geq0$ is the \emph{global amplitude} of the tone,
$D\in[0,1]^{N_{\mathrm{har}}\times N_{\mathrm{ins}}}$ is the \emph{dictionary}
containing the \emph{relative amplitudes} for the harmonics,
and $\tilde{\varphi}_h\in[-\pi,\pi)$ is the phase angle for the respective harmonic.

In a realistic music recording, the tones of the different instruments
overlap and their parameters change over time.  Thus, to construct an
appropriate time-frequency model, we must equip the parameters with indices
relating to the tones $j=1,\dotsc,m$ (where $m\in\N$
is the total number of simultaneously played tones)
and the time frames $k=1,\dotsc,n_{\mathrm{len}}$.  We define the tone-wise
and global model spectrograms, respectively, as:%
\begin{subequations}\label{eq:specmodel}
\begin{equation}
  \begin{aligned}
  z_j[k,l]&=\sum_{h} a_{j,h,k}\exp\Biggl(-\frac{(\beta l-f_{j,h,k})^2}{2\sigma_{j,k}^2}+i2\pi f_{j,h,k}\alpha k\Biggr)\\
  &=\sum_{h} a_{j,k}\,D[h,\eta_{j,k}]\exp\Biggl(-\frac{(\beta l-f_{j,h,k})^2}{2\sigma_{j,k}^2}+i\varphi_{j,h,k}\Biggr)
  \end{aligned}
\end{equation}
incorporating the phase shift via
$\varphi_{j,h,k}\coloneqq\tilde{\varphi}_{j,h,k}+2\pi f_{j,h,k}\alpha k$, and:
\begin{equation}
  z[k,l]=\sum_{j} z_j[k,l].
\end{equation}
\end{subequations}
For signals consisting only of sinusoids,
the spectrum is modeled precisely by
\eqref{eq:tonespec}, where the standard deviation $\sigma_{j,k}$ of the
Gaussians is given as $\sigma_{j,k}=1/(2\pi\zeta)$.
However, especially at the beginning and end of a tone, boundary effects can
occur, leading to the spectrum being better approximated by different values
for $\sigma_{j,k}$. Therefore we include those as free parameters.

\section{Learned Separation}

\subsection{Distance Function}

The parametrized model $z[k,l]$ from \eqref{eq:specmodel} should match the
time-frequency representation $Z[k,l]$ from \eqref{eq:sample} as closely
as possible.  To formalize this, we need to define a distance function.
While the $\ell_2$ distance would be simple and \enquote{canonical,}
the problem is that it overemphasizes the correctness of the
high-volume parts of the representation rather than the structural
similarity.  A common alternative is the $\beta$-divergence
\cite[cf.][]{Fevotte11}
(which is a generalization of the $\ell_2$ distance, the Kullback-Leibler
divergence, and the Itakura-Saito divergence), but this leads to problems
with unexplained noise in the spectrum.  Instead, we use the distance
measure introduced in \cite{Schulze21}; for a given time frame $k$, we
set $y=z[k,\cdot]$ and $Y=Z[k,\cdot]$ and define:
\begin{equation}
d_{2,\delta}^{q,\mathrm{abs}}(Y,y)=\frac{1}{2}\sum_l
\Bigl(\bigl(\sabs{Y[l]}+\delta\bigr)^q-\bigl(\sabs{y[l]}+\delta\bigr)^q\Bigr)^2,\qquad
\quad q\in(0,1],\quad\delta>0.
\label{eq:liftdist}
\end{equation}
The $q$ exponent has the purpose of \emph{lifting} the low-volume parts
of the representation in order to increase their relevance.  The canonical
choice is $q=1/2$, since this is the lowest value to keep the expression
convex in $y$.  The value for $\delta$ can be low, as it is merely there
to ensure differentiability at $y[l]=0$.
Unlike the $\beta$-divergence, the distance function in \eqref{eq:liftdist}
is symmetric, but it is still not a metric in the mathematical sense.

\subsection{Model Fitting}

Even though \eqref{eq:liftdist} is convex in $y$ for $q=1/2$,
the spectrum $y$ itself is not point-wise globally convex
in all the parameters appearing in \eqref{eq:specmodel}.
Therefore, conventional optimization
methods based on gradient descent are not a good choice for minimizing
$d_{2,\delta}^{q,\mathrm{abs}}(Y,y)$.
Instead, we use a deep neural network to predict the tone parameters.
While some of these parameters (the \emph{deterministic parameters})
can be trained normally via backpropagation, we treat the
\enquote{problematic} parameters as \emph{stochastic parameters}
which are trained via
\emph{policy gradients} \cite{Williams92}.

The neural network is applied in such the way that it predicts
the parameters for one tone at a time.  After each prediction,
the spectrum for that tone is computed and provided back to the
network for the following steps.

\subsubsection{Parameter Representation}

We first have to decide which parameters are stochastic and which ones
are deterministic.  The fundamental frequency parameter $f_{j,1,k}^\circ$
is clearly one of those in which $y$ is not convex; however, it does
have a useful gradient on a local scale.
We therefore split this parameter into two parameters as
$f_{j,1,k}^\circ=\beta(\nu_{j,k}+\tilde{\nu}_{j,k})$, where the values for
$\nu_{j,k}\in\N$ are discrete and those for $\tilde{\nu}_{j,k}\in\R$ are
continuous.  We treat
$\nu_{j,k}$ as a stochastic parameter and $\tilde{\nu}_{j,k}$
as a deterministic one.

If we limit each instrument to exactly one tone (therefore,
$m=N_{\mathrm{ins}}$), the instrument parameter
$\eta_{j,k}$ is not required mathematically.  However, it makes
sense to include it from a practical algorithmic perspective, since this
allows for a network architecture that sequentially extracts one tone after
another while freely choosing the extraction order (see Section
\ref{sec:policy_grads}).  As $\eta_{j,k}$ is discrete, it is impossible to
obtain a gradient.  Thus, we model it as a stochastic parameter following a
categorical distribution.

For $\sigma_{j,k}$, while it is possible
to find examples in which the distance $d_{2,\delta}^{q,\mathrm{abs}}(Y,y)$
is not convex with respect to it,
the gradient around the theoretical value is usually good,
so we can treat it as a deterministic parameter.

The inharmonicity parameter $b_{j,k}$ is also continuous and has a good
gradient around the optimum, but depending on the characteristics of the
instrument, there can exist local optima.  Therefore we do not rely
on backpropagation and instead treat it as a stochastic parameter
following a gamma distribution (see Figure~\ref{fig:gamma_dist}).
The reason why we chose a gamma distribution is that it is non-negative
and it can have two qualitatively different shapes:  Either it tends
towards $\infty$ at zero (which is useful to model tones without
inharmonicity), or it is bell-shaped around a finite maximum
(to model tones with inharmonicity).  Approaching infinity, it always decays
exponentially, and in the edge case between the two shapes, it matches
an exponential distribution.

In the tone amplitudes $a_{j,k}$, the problem is convex, and therefore
they are treated as deterministic parameters.  In total,
the stochastic parameters for each tone are
$\varpi_{\mathrm{s},j,k}=(\nu_{j,k},\eta_{j,k},b_{j,k})$, and the
deterministic parameters are
$\varpi_{\mathrm{d},j,k}=(a_{j,k},\tilde{\nu}_{j,k},\sigma_{j,k})$.
When it is clear which time frame we are considering, we can drop the
dependency on $k$ and summarize:%
\begin{subequations}
\begin{gather}
  \varpi_{\mathrm{s}}=(\varpi_{\mathrm{s},1},\dotsc,\varpi_{\mathrm{s},m})
  =(\varpi_{\mathrm{s},1,k},\dotsc,\varpi_{\mathrm{s},m,k}),\\
  \varpi_{\mathrm{d}}=(\varpi_{\mathrm{d},1},\dotsc,\varpi_{\mathrm{d},m})
  =(\varpi_{\mathrm{d},1,k},\dotsc,\varpi_{\mathrm{d},m,k}).
\end{gather}
\end{subequations}

\begin{figure}
    \small\centering
    \tikzsetnextfilename{Figure_\the\numexpr(\thefigure+1)\relax}
    \begin{tikzpicture}
        \begin{axis}[width=0.7\textwidth,height=7cm,enlargelimits=false,
                legend cell align=left,
                xlabel=,ylabel={Probability density of $\Gamma(\alpha^\Gamma, \beta^\Gamma)$}, ymax=0.62,
                y tick label style={/pgf/number format/.cd},
                xshift=0.25\textwidth]
            \addplot[black, solid, thick] table
                    [x expr=\thisrowno{0},
                    y expr=\thisrowno{1}] {plot_gamma_dist.dat};
            \addlegendentry{$\alpha^\Gamma=\eqmakebox[alpha][l]{0.5,}\;\beta^\Gamma=1$}
            \addplot[blue, dashed, thick] table
                    [x expr=\thisrowno{0},
                    y expr=\thisrowno{2}] {plot_gamma_dist.dat};
            \addlegendentry{$\alpha^\Gamma=\eqmakebox[alpha][l]{1,}\;\beta^\Gamma=1$}
            \addplot[red, loosely dashed, thick] table
                    [x expr=\thisrowno{0},
                    y expr=\thisrowno{3}] {plot_gamma_dist.dat};
            \addlegendentry{$\alpha^\Gamma=\eqmakebox[alpha][l]{1,}\;\beta^\Gamma=0.5$}
            \addplot[green!60!black, densely dashdotted, thick] table
                    [x expr=\thisrowno{0},
                    y expr=\thisrowno{4}] {plot_gamma_dist.dat};
            \addlegendentry{$\alpha^\Gamma=\eqmakebox[alpha][l]{1.5,}\;\beta^\Gamma=0.25$}
            \addplot[orange!90!black, densely dotted, thick] table
                    [x expr=\thisrowno{0},
                    y expr=\thisrowno{5}] {plot_gamma_dist.dat};
            \addlegendentry{$\alpha^\Gamma=\eqmakebox[alpha][l]{3,}\;\beta^\Gamma=1$}
            \addplot[violet!90!white, dotted, thick] table
                    [x expr=\thisrowno{0},
                    y expr=\thisrowno{6}] {plot_gamma_dist.dat};
            \addlegendentry{$\alpha^\Gamma=\eqmakebox[alpha][l]{3,}\;\beta^\Gamma=0.5$}
        \end{axis}
    \end{tikzpicture}
    \caption{Probability density functions of gamma distributions for different parameter choices. For $\alpha^\Gamma \geq 1$, the function has the mode at $(\alpha^\Gamma-1)/\beta^\Gamma$, while for $\alpha^\Gamma < 1$, the function tends to $\infty$ at zero.  In our proposed method, the network selects a distribution shape for each inharmonicity coefficient $b_j$ by its outputs $\alpha^\Gamma_j,\beta^\Gamma_j$.}
    \label{fig:gamma_dist}
\end{figure}
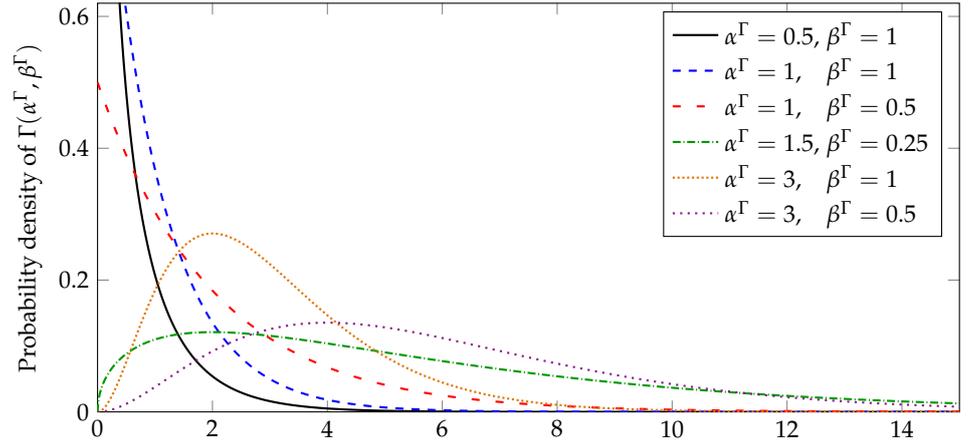

\subsubsection{Policy Gradients}
\label{sec:policy_grads}

We use a neural network with network parameters $\theta$ both
to give the \emph{policy}
$\pi_\theta(\varpi_{\mathrm{s}}|Y)$, which is the
(discrete or continuous) probability density for
the stochastic parameters $\varpi_{\mathrm{s}}$
given the input spectrum $Y$ and also to compute the deterministic
parameters $\varpi_{\mathrm{d}}$ from the input and the stochastic parameters.
With this, we can express the \emph{loss function} as:
\begin{equation}
  L(\varpi_{\mathrm{s}},\theta,D,Y)\coloneqq
  L\bigl(\varpi_{\mathrm{s}},\varpi_{\mathrm{d}}(\varpi_{\mathrm{s}},\theta,Y),D,Y\bigr)
  =d_{2,\delta}^{q,\mathrm{abs}}\bigl(Y,y(\varpi_{\mathrm{s}},\varpi_{\mathrm{d}},D)\bigr)
  =d_{2,\delta}^{q,\mathrm{abs}}(Y,y).
\end{equation}

For the expected loss, we follow the usual computation \cite[cf.][]{Sutton20}
but also apply the product rule for the deterministic parameters:
\begin{equation}\label{eq:polgrad}
\begin{aligned}
  &\nabla_\theta\expect_{\pi_{\theta}(\varpi_\mathrm{s}|Y)}\bigl[L(\varpi_\mathrm{s},\theta,D,Y)\bigr]\\
  &=\nabla_\theta\int \pi_{\theta}(\varpi_{\mathrm{s}}|Y)\,L(\varpi_{\mathrm{s}},\theta,D,Y)\:\dif\varpi_{\mathrm{s}}\\
  &=\int \nabla_\theta\pi_{\theta}(\varpi_{\mathrm{s}}|Y)\,L(\varpi_{\mathrm{s}},\theta,D,Y)+\pi_{\theta}(\varpi_{\mathrm{s}}|Y)\,\nabla_\theta L(\varpi_{\mathrm{s}},\theta,D,Y)\:\dif\varpi_{\mathrm{s}}\\
  &=\int \pi_{\theta}(\varpi_{\mathrm{s}}|Y)\,\biggl(\frac{\nabla_\theta\pi_{\theta}(\varpi_{\mathrm{s}}|Y)}{\pi_{\theta}(\varpi_{\mathrm{s}}|Y)}\,L(\varpi_{\mathrm{s}},\theta,D,Y)+\nabla_\theta L(\varpi_{\mathrm{s}},\theta,D,Y)\biggr)\:\dif\varpi_{\mathrm{s}}\\
  &=\expect_{\pi_{\theta}(\varpi_{\mathrm{s}}|Y)}\bigl[\nabla_\theta\log\pi_{\theta}(\varpi_{\mathrm{s}}|Y)\,L(\varpi_{\mathrm{s}},\theta,D,Y)+\nabla_\theta L(\varpi_{\mathrm{s}},\theta,D,Y)\bigr],
\end{aligned}
\end{equation}
where we refer to the first term in the sum as the \emph{policy gradient}
and on the second term as the \emph{backpropagation gradient}.
The applicability of the Leibniz integral rule can be shown under
realistic conditions.

The total dimensionality of $\varpi_{\mathrm{s}}$ is too high to
represent $\pi_{\theta}(\varpi_{\mathrm{s}}|Y)$ as a whole.
Therefore, we decompose
the log-probability density into:
\begin{equation}
  \begin{aligned}
    &\log\pi_{\theta}(\varpi_{\mathrm{s}}|Y)\\
    &=\log\bigl(\pi_{\theta}(\varpi_{\mathrm{s},1}|Y)\cdot\pi_{\theta}(\varpi_{\mathrm{s},2}|Y,\varpi_{\mathrm{s},1})\cdots\pi_{\theta}(\varpi_{\mathrm{s},m}|Y,\varpi_{\mathrm{s},1},\dotsc,\varpi_{\mathrm{s},m-1})\bigr)\\
    &=\log\pi_{\theta}(\varpi_{\mathrm{s},1}|Y)+\log\pi_{\theta}(\varpi_{\mathrm{s},2}|Y,\varpi_{\mathrm{s},1})+\ldots+\log\pi_{\theta}(\varpi_{\mathrm{s},m}|Y,\varpi_{\mathrm{s},1},\dotsc,\varpi_{\mathrm{s},m-1}).
  \end{aligned}
\end{equation}

In practice, we
first sample the stochastic parameters $\varpi_{\mathrm{s},1}$ for the
first tone, then obtain the deterministic parameters $\varpi_{\mathrm{d},1}$,
and from these compute a model spectrum $y_1$.  For each additional tone
$j=2,\dotsc,m$,
we gain $\varpi_{\mathrm{s},j}$ and $\varpi_{\mathrm{d},j}$
depending on the previous spectra $y_1,\dotsc,y_{j-1}$, which encapsulate
all the relevant information about the parameters of the previous tones.

The parameters $\nu_j,\eta_j$ are sampled via a joint categorical distribution,
making sure that no instrument plays more than one tone.  For each
possible value of $(\nu_j,\eta_j)$, the network gives a value for the
deterministic parameters $\varpi_{\mathrm{d},j}$ as well as for the parameters
$\alpha_j^\Gamma,\beta_j^\Gamma$ for the gamma distribution for $b_j$.

\subsection{Phase Prediction}

Of the tone parameters for \eqref{eq:dictmodel}, we still need the
phase angles $\varphi_{j,h}$.  Canonically, we could represent them as
a vector that is output by the network for each possible choice of
$\nu_j,\eta_j$, but this would lead to high dimensionality.
Instead, we let the network emit a single artificial spectrum
$v_j\in\C^{n_{\mathrm{spc}}}$
for each possible instrument choice $\eta_j$.  This is used as the right-hand
side of a least-squares problem for determining the coefficients $c_{j,h}\in\C$:
\begin{equation}\label{eq:coeffsys}
\min_{(c_{j,h})}\frac{1}{2}\sum_{l}\Biggabs{\sum_h c_{j,h}\cdot
\exp\Biggl(-\frac{\Bigl(\beta
  l-f_{j,1}^\circ h\sqrt{1+b_{j}h^2}\Bigr)^2}{2\sigma_{j}^2}\Biggr) - v_j[l]}^2,\qquad
l=0,\dotsc,n_{\mathrm{spc}}-1,
\end{equation}
from which we extract the phase angles as $\varphi_{j,h}=\arg c_{j,h}$.
We apply some $\ell_2$ regularization to improve the condition number
of the system.
For $c_{j,h}=0$, the phase would be ill-defined,
but since the magnitudes
of these coefficients are stabilized via the objectives introduced in
Section \ref{sec:cmplxobj}, we do not realistically expect this case to
happen.

With this approach, the frequency dimension of the network output is used
differently than for the other tone parameters.
Instead of providing an output for each possible tone choice $\nu_j$, the
network computes a single spectrum, which then determines the phases of all
harmonics.  We expect the spatial structure of the U-Net architecture with
respect to the frequency dimension to be beneficial for this task.
While $c_{j,h}$ is not explicitly given as a conditional value
depending on $\varpi_{\mathrm{s}}$, the computation of
$c_{j,h}$ in \eqref{eq:coeffsys}
does depend on all the stochastic parameters
(even $b_j$), and since typically $n_{\mathrm{spc}}>N_{\mathrm{har}}$,
the network has some freedom to output $v_j$ such that $c_{j,h}$ take
different values depending on the other parameters.

Since $v_j$ is the right-hand side of a linear least-squares system,
optimizing $y$ with respect to $v_j$ is convex.  Therefore, training
is done deterministically via backpropagation through the pseudo-inverse.
Also, we include the additional gradient
that occurs with respect to the left-hand-side parameters that appear inside
the exponential.\footnote{%
  Computation of the gradient of the solution of a least-squares system
  with respect to the left-hand side is usually included in automatic
  differentiation frameworks, but to obtain it explicitly, one can
  repeatedly apply the Woodbury formula.}

In the case where the positions of the peaks from the model perfectly match
those from the spectrum $Y$ without any overlap or additive noise, the choice
$v_j=Y$ gives the ideal phase values $\varphi_{j,h}$ for all
$j=1,\dotsc,m$ and $h=1,\dotsc,N_{\mathrm{har}}$.  While this exact case is
not realistic (not least since Gaussians have unbounded support),
a network with skip connections can quickly learn to predict good approximate
phase values.

\subsection{Complex Objectives}
\label{sec:cmplxobj}

So far, we have only used the parameters $c_{j,h}$ from \eqref{eq:coeffsys}
in order to determine the phase values $\varphi_{j,h}$.  However, we can
also use them for a different purpose:  While the dictionary representation
\eqref{eq:dictmodel} is necessary in order to distinguish the instruments,
it is never fully accurate since the relation of the amplitudes
of the harmonics can vary slightly even for the same instrument.
The typical remedy for this is \emph{spectral masking}, but as
explained in \cite{Schulze21}, this process does not properly deal with
interference.  Instead, we create a
\emph{direct prediction}, in which we replace the tone amplitudes and
dictionary entries in \eqref{eq:specmodel}
with the parameters $c_{j,h,k}$ (with time-frame dependency added back in):%
\begin{subequations}\label{eq:dirpred}
  \begin{align}
    z_j^{\mathrm{dir}}[k,l]&=\sum_{h} c_{j,h,k}\exp\Biggl(-\frac{(\beta l-f_{j,h,k})^2}{2\sigma_{j,k}^2}\Biggr),\\
    z^{\mathrm{dir}}[k,l]&=\sum_{j} z_j^{\mathrm{dir}}[k,l].
\end{align}
\end{subequations}
For a fixed $k$, we set $y^{\mathrm{dir}}=z^{\mathrm{dir}}[k,\cdot]$.

The distance function from \eqref{eq:liftdist} only considers the absolute
value and ignores the phase entirely.  This is not necessarily a problem,
but knowing the exact phase angles would be useful for resynthesis, and
it makes more sense as a training objective.  Therefore, we introduce a
modified distance function that respects the phase:
\begin{equation}
d_{2,\delta}^{q,\mathrm{rad}}(Y,y)=\frac{1}{2}\sum_l
\biggabs{\bigl(\sabs{Y[l]}+\delta\bigr)^q\cdot\frac{Y[l]}{\sabs{Y[l]}}-\bigl(\sabs{y[l]}+\delta\bigr)^q\cdot\frac{y[l]}{\sabs{y[l]}}}^2.
\label{eq:liftcmplx}
\end{equation}
To avoid division by zero, we add a tiny positive constant to the denominators.

\end{paracol}
\nointerlineskip
\begin{figure}
  \widefigure
  \small\centering
  \tikzsetnextfilename{Figure_\the\numexpr(\thefigure+1)\relax}
  \begin{tikzpicture}
    \begin{groupplot}[group style={group size=2 by 3,horizontal sep=0.1\textwidth,vertical sep=3.75em},
                      width=0.48\textwidth,title style={yshift=-1ex}]
      \nextgroupplot[title={Direct prediction $y^{\mathrm{dir}}$ $=$ $Y$},height=5.5cm,enlargelimits=false,
            legend pos=north east,legend cell align=left,
            xlabel=$l$,ylabel={Absolute value},ymax=0.225]
        \addplot[red, solid, thick] table
                [x expr=\thisrowno{0},
                y expr=\thisrowno{3}] {plot_interference.dat};
        \addlegendentry{Peak 1}
        \addplot[blue, solid, thick] table
                [x expr=\thisrowno{0},
                y expr=\thisrowno{7}] {plot_interference.dat};
        \addlegendentry{Peak 2}
        \addplot[black, dashed, thick] table
                [x expr=\thisrowno{0},
                y expr=\thisrowno{11}] {plot_interference.dat};
        \addlegendentry{Sum}
      \nextgroupplot[title={Dictionary-based prediction $y$},height=5.5cm,enlargelimits=false,
            xlabel=$l$,ylabel={Absolute value}, ymax=0.225]
        \addplot[red, solid, thick] table
                [x expr=\thisrowno{0},
                y expr=\thisrowno{15}] {plot_interference.dat};
        \addplot[blue, solid, thick] table
                [x expr=\thisrowno{0},
                y expr=\thisrowno{19}] {plot_interference.dat};
        \addplot[black, dashed, thick] table
                [x expr=\thisrowno{0},
                y expr=\thisrowno{23}] {plot_interference.dat};
      \nextgroupplot[height=3.25cm,enlargelimits=false,
            xlabel=$l$,ylabel={Phase}, ymin=-0.3, ymax=1.8708,
            ytick={0, 0.7854, 1.5708}, yticklabels={$0$, $\pi/4$, $\pi/2$}]
        \addplot[red, solid, thick] table
                [x expr=\thisrowno{0},
                y expr=\thisrowno{4}] {plot_interference.dat};
        \addplot[blue, solid, thick] table
                [x expr=\thisrowno{0},
                y expr=\thisrowno{8}] {plot_interference.dat};
        \addplot[black, dashed, thick] table
                [x expr=\thisrowno{0},
                y expr=\thisrowno{12}] {plot_interference.dat};
      \nextgroupplot[height=3.25cm,enlargelimits=false,
            xlabel=$l$,ylabel={Phase}, ymin=-0.3, ymax=1.8708,
            ytick={0, 0.7854, 1.5708}, yticklabels={$0$, $\pi/4$, $\pi/2$}]
        \addplot[red, solid, thick] table
                [x expr=\thisrowno{0},
                y expr=\thisrowno{16}] {plot_interference.dat};
        \addplot[blue, solid, thick] table
                [x expr=\thisrowno{0},
                y expr=\thisrowno{20}] {plot_interference.dat};
        \addplot[black, dashed, thick] table
                [x expr=\thisrowno{0},
                y expr=\thisrowno{24}] {plot_interference.dat};
    \nextgroupplot[group/empty plot,height=4.25cm]
    \nextgroupplot[height=4.25cm,enlargelimits=false,
            legend pos=north east,legend cell align=left,
            xlabel=$l$,ylabel={Loss contribution}, ymax=0.11,
            y tick label style={/pgf/number format/.cd, fixed, fixed zerofill}]
        \addplot[solid, thick] table
                [x expr=\thisrowno{0},
                y expr=\thisrowno{25}] {plot_interference.dat};
        \addlegendentry{$d^{q,\mathrm{abs}}_{2,\delta}(Y,y)$}
        \addplot[dotted, thick] table
                [x expr=\thisrowno{0},
                y expr=\thisrowno{26}] {plot_interference.dat};
        \addlegendentry{$d^{q,\mathrm{rad}}_{2,\delta}(Y,y)$}
    \end{groupplot}
  \end{tikzpicture}
  \caption{Example showing the interference of two overlapping peaks. Each peak models the contribution of one harmonic of a tone to the spectrum (cf.~\eqref{eq:tonespec}). The left plots show a part of a direct prediction $y^{\mathrm{dir}}$, which is assumed to equal the true spectrum $Y$ for this example, and the right plots show a dictionary-based prediction $y$ with deviating amplitudes. Due to the different amplitudes, also the phases mix differently, leading to a high value of $d^{q,\mathrm{rad}}_{2,\delta}(Y,y)$. The phases could be optimized for $y$ (by increasing the phase for peak 1 and/or peak 2), but this would lead to suboptimal phases in $y^{\mathrm{dir}}$. In contrast, the used loss $d^{q,\mathrm{abs}}_{2,\delta}(Y,y)$ does not depend on the phase.}
  \label{fig:example_interference}
\end{figure}
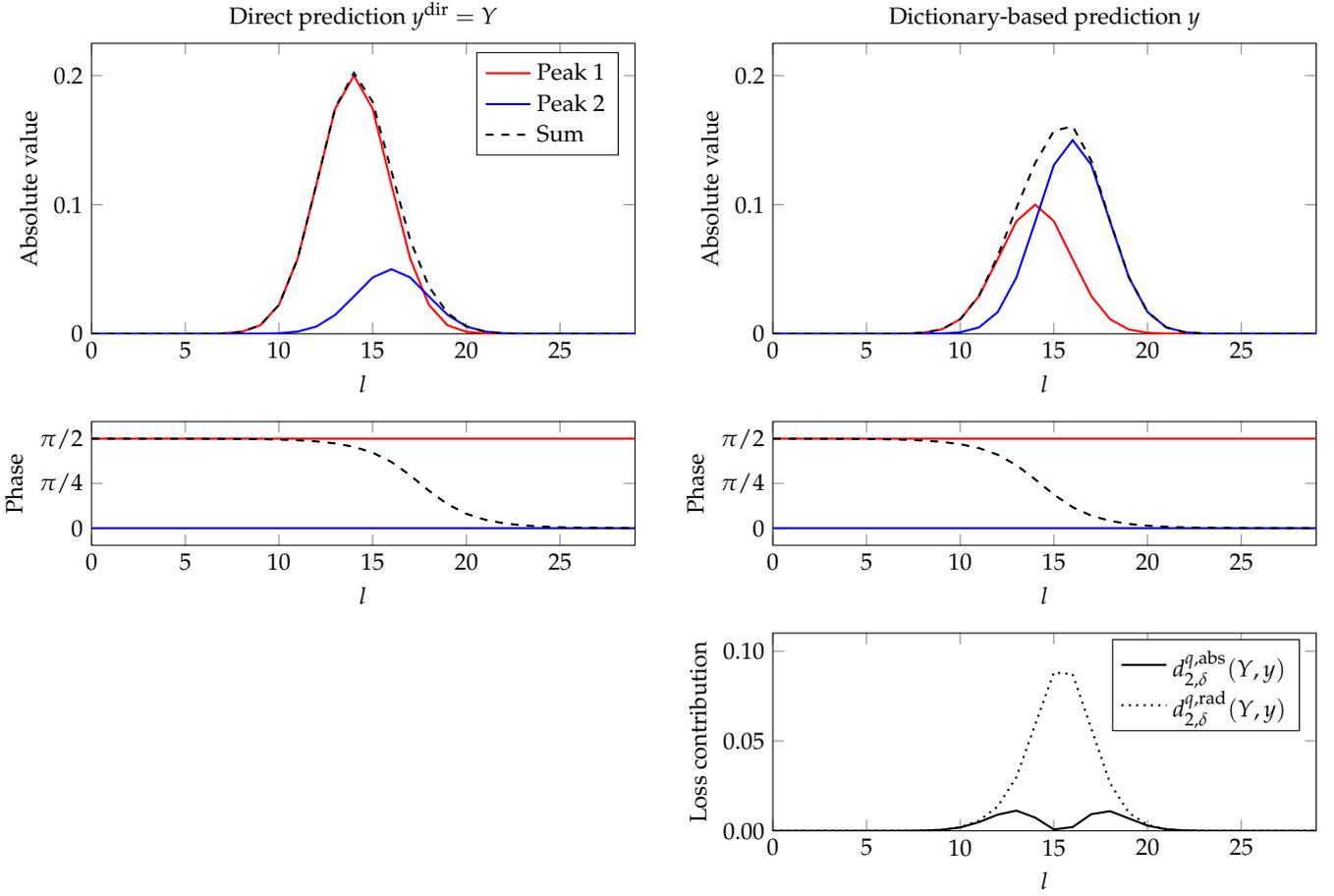
\begin{paracol}{2}
\makeatletter\ifthenelse{\equal{\@status}{submit}}{\linenumbers}{}\makeatother
\switchcolumn

\end{paracol}
\nointerlineskip
\begin{figure}
\widefigure
\small\centering
\tikzsetnextfilename{Figure_\the\numexpr(\thefigure+1)\relax}
\begin{tikzpicture}
    \begin{groupplot}[group style={group size=2 by 1,horizontal sep=0.1\textwidth,vertical sep=3.25em},
                    width=0.48\textwidth,title style={yshift=-1ex}]
    \nextgroupplot[title={Direct prediction $y^{\mathrm{dir}}$},height=5.5cm,enlargelimits=false,
            legend pos=north east,legend cell align=left,
            xlabel=$l$,ylabel={Real value},ymin=-0.025,ymax=0.225]
        \addplot[red, solid, thick] table
                [x expr=\thisrowno{0},
                y expr=\thisrowno{1}] {plot_regularization.dat};
        \addlegendentry{$y^{\mathrm{dir}}_1$}
        \addplot[blue, solid, thick] table
                [x expr=\thisrowno{0},
                y expr=\thisrowno{2}] {plot_regularization.dat};
        \addlegendentry{$y^{\mathrm{dir}}_2$}
        \addplot[black, dashed, thick] table
                [x expr=\thisrowno{0},
                y expr=\thisrowno{3}] {plot_regularization.dat};
        \addlegendentry{Sum}
    \nextgroupplot[title={Dictionary-based prediction $y$},height=5.5cm,enlargelimits=false,
            legend pos=north east,legend cell align=left,
            xlabel=$l$,ylabel={Real value},ymin=-0.025,ymax=0.225]
        \addplot[red, solid, thick] table
                [x expr=\thisrowno{0},
                y expr=\thisrowno{4}] {plot_regularization.dat};
        \addlegendentry{$y_1$}
        \addplot[blue, solid, thick] table
                [x expr=\thisrowno{0},
                y expr=\thisrowno{5}] {plot_regularization.dat};
        \addlegendentry{$y_2$}
        \addplot[black, dashed, thick] table
                [x expr=\thisrowno{0},
                y expr=\thisrowno{6}] {plot_regularization.dat};
        \addlegendentry{Sum}
    \end{groupplot}
\end{tikzpicture}
\caption{Example showing non-uniqueness of the tone separation in the direct prediction. In the direct prediction (left) the separation of the two instruments is different from the dictionary-based prediction $y$ (right). For this example, we assume the dictionary-based separation to be correct, so ideally the direct separation would be the same. However, the total spectrum (Sum) of the incorrect separation equals the true total spectrum and thus also achieves the optimal loss value $d_{2,\delta}^{q,\mathrm{rad}}(Y,y^{\mathrm{dir}})$. This motivates to regularize the individual tones $y^{\mathrm{dir}}_j$ of the direct prediction using $y_j$. The imaginary parts of the spectra are assumed to be all zero for this example.}
\label{fig:example_regularization}
\end{figure}
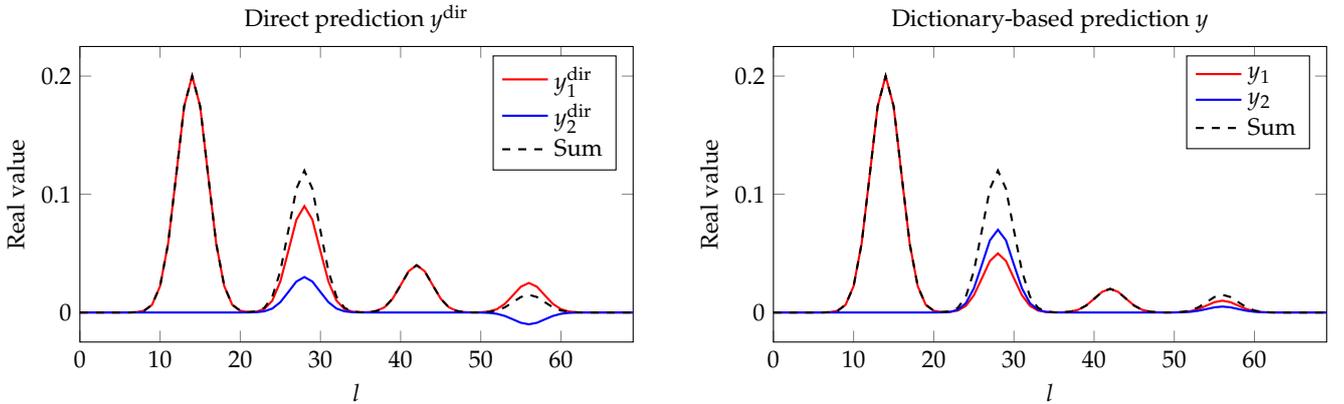
\begin{paracol}{2}
\makeatletter\ifthenelse{\equal{\@status}{submit}}{\linenumbers}{}\makeatother
\switchcolumn

We can now compare $y$ to $Y$ (as before), $y^{\mathrm{dir}}$ to $Y$, and
$y$ to $y^{\mathrm{dir}}$, and we can choose between the distance functions
$d_{2,\delta}^{q,\mathrm{abs}}$ and $d_{2,\delta}^{q,\mathrm{rad}}$
as defined in \eqref{eq:liftdist}, \eqref{eq:liftcmplx}.  While it would
be ideal to have the phases matching in all spectra, there are some
caveats:
\begin{itemize}
\item It takes a number of training iterations for $v_j$ to give a useful
  value.  In the meantime, the training of the other parameters can go
  in a bad direction.
\item If the discrepancy between $y$ and $y^{\mathrm{dir}}$ is high
  \emph{and} there is a lot of overlap between the peaks
  (typically from different tones),
  the optimal phase values for $y$ and $y^{\mathrm{dir}}$
  may be significantly different.  An example for this is displayed in
  Figure~\ref{fig:example_interference}:  The two peaks (red and blue) each
  have different phases, but by design, those are identical between the
  predictions.  However, since the dictionary prediction is less flexible,
  its amplitude magnitudes of the harmonics often do not accurately match the
  input spectrum $Y$, which shifts the phase in the overlapping region.
  Thus, attempting to minimize both
  $d_{2,\delta}^{q,\mathrm{rad}}(Y,y^{\mathrm{dir}})$ and
  $d_{2,\delta}^{q,\mathrm{rad}}(Y,y)$ would lead to a conflict regarding
  the choice of common phase values.
\end{itemize}
Therefore, we continue to compare $y$ and $Y$ via
$d_{2,\delta}^{q,\mathrm{abs}}$ (without the phase),
and we use $d_{2,\delta}^{q,\mathrm{rad}}$ for the comparison between
$y^{\mathrm{dir}}$ and $Y$, since $y^{\mathrm{dir}}$ is the spectrum that we
will end up using for resynthesis, so we aim for the phase to be correct.
Between $y$ and $y^{\mathrm{dir}}$,
we can compare tone-wise, but since some discrepancy is to be expected,
we only associate it with a small penalty; the purpose is to regularize
$y^{\mathrm{dir}}$ for the case that the peaks of different tones overlap
so much that the $c_{j,h}$ are not unique (see
Figure~\ref{fig:example_regularization}).
For this task, we employ the loss terms
$d_{2,\delta}^{q,\mathrm{rad}}(y^{\mathrm{dir}}_j,y_j)$ to compare the tone
spectra $y_j^{\mathrm{dir}}\coloneqq z_j^{\mathrm{dir}}[k,\cdot]$
and $y_j\coloneqq z_j[k,\cdot]$ in both magnitude and phase.
However, since the individual peaks making up $y^{\mathrm{dir}}_j$ and $y_j$
necessarily have the same phases, the difference between
$d_{2,\delta}^{q,\mathrm{rad}}$ and $d_{2,\delta}^{q,\mathrm{abs}}$ only matters
if there is significant overlap between the peaks within the same tone,
which is the case at very low fundamental frequencies.

The loss functions $d_{2,\delta}^{q,\mathrm{rad}}$ and $d_{2,\delta}^{q,\mathrm{abs}}$
are both based on the $\ell_2$ loss, so they do not induce sparsity.
Thus, if there is linear dependency in the dictionary or, more likely,
there is a discrepancy between the dictionary model for an instrument
and an actual tone played by that instrument, then one single tone played
by an instrument may get identified as multiple ones, either with the same
fundamental frequency or with overlapping harmonics.
Thus, we introduce
an additional \emph{sparsity parameter} $u_j\in\{0,1\}$ that indicates
whether a certain tone is present at all, and we henceforth include it in
the set of stochastic parameters $\varpi_{\mathrm{s},j}$.
Whenever $u_j=0$, we discard the tone in the \emph{sparse prediction}
given by
\begin{align}
y^{\mathrm{spr}}&=\sum_{j} u_j\,y_j
\end{align}
and discount part of the loss.
In terms of architecture, the parameter $u_j$ is modeled as a Bernoulli
distribution. %

With $y_j^{\mathrm{dir}}=z_j^{\mathrm{dir}}[k,\cdot]$, $y_j=z_j[k,\cdot]$,
and $y^{\mathrm{dir}}=z^{\mathrm{dir}}[k,\cdot]$ again, we define the loss as:
\begin{equation}
L(\varpi_{\mathrm{s}},\theta,D,Y)
=\mu_1\,d_{2,\delta}^{q,\mathrm{abs}}(Y,y^{\mathrm{spr}})\cdot\lambda^{\sum_j(1-u_j)}
+\mu_2\,d_{2,\delta}^{q,\mathrm{rad}}(Y,y^{\mathrm{dir}})
+\frac{\mu_3}{m}\sum_jd_{2,\delta}^{q,\mathrm{rad}}(y^{\mathrm{dir}}_j,y_j),
\end{equation}
choosing $\lambda=0.9$.  This value is somewhat arbitrary, so we do not
use it to enforce sparsity directly in $y$.  Instead, we compute the
loss in the first term based on $y^{\mathrm{spr}}$ so that the parameters
are \emph{compatible} with a sparse solution and do not rely on redundant tones.
However, additional tones can still appear in $y^{\mathrm{dir}}$ to reduce the
distance to $Y$ and, by extension,
also in $y$ to reduce the distance to $y^{\mathrm{dir}}$.

We give both distances to $Y$ the same loss coefficients while the
regularization is supposed to be small.  In practice, we find that
$\mu_1=10$, $\mu_2=10$, $\mu_3=1$ is a good choice.

\subsection{Sampling for Gradient Estimation}

In order to apply a gradient descent method, we need to compute
the expectation in \eqref{eq:polgrad}.  However, the set of possible
parameters $\varpi_{\mathrm{s}}$ is much too large to do so analytically,
so we have to estimate it instead.  For this, we use:
\begin{equation}\label{eq:est}
\hat{g}_{\pi_{\theta},Y}=\frac{1}{S}\sum_{i=1}^S\Bigl(\nabla_\theta\log\pi_{\theta}(\varpi_{\mathrm{s}}^i|Y)\cdot \bigl(L(\varpi_{\mathrm{s}}^i,\theta,D,Y)-C(\theta,D,Y)\bigl)+\nabla_\theta L(\varpi_{\mathrm{s}}^i,\theta,D,Y)\Bigr),
\end{equation}
with $\varpi_{\mathrm{s}}^1,\dotsc,\varpi_{\mathrm{s}}^S\sim\pi_{\theta}(\varpi_{\mathrm{s}}|Y)$,
where $S\in\N$ is the number of samples and $C(\theta,D,Y)$ is the
\emph{baseline} \cite[cf.][]{Sutton20}.
The baseline plays a crucial role in reducing the
variance of the gradient, and a common choice is:
\begin{equation}
C(\theta,D,Y)=\expect_{\pi_{\theta}(\varpi_{\mathrm{s}}|Y)}\bigl[L(\varpi_{\mathrm{s}},\theta,D,Y)\bigr].
\end{equation}
As long as the baseline is independent of the samples
$\varpi_{\mathrm{s}}^1,\dotsc,\varpi_{\mathrm{s}}^S$, the estimator
\eqref{eq:est} is unbiased.  However, the easiest way to
estimate the baseline is via:
\begin{equation}\label{eq:blest}
\hat{C}(\theta,D,Y)=\frac{1}{S}\sum_{i=1}^SL(\varpi^i_{\mathrm{s}},\theta,D,Y),
\end{equation}
which obviously depends on the samples and therefore introduces
a bias factor of $(S-1)/S$ for the policy gradient.  If desired,
one could correct for this bias by dividing by this factor, but
it is generally advisable to be conservative about the policy gradient,
so we choose to leave it uncorrected.  However, the value for $S$
should be chosen large enough for \eqref{eq:blest} to be a good estimator.

During the training, it is expected that the policy will become increasingly
deterministic, so less exploration will be performed.  The problem with this
is that multiple parameters need to be trained, and it is easy for the
training process to get stuck in local minima.  Therefore, we should
encourage the exploration of more parameter choices even when the policy
has stabilized.  Inspired by the algorithm from \cite{Silver17},
we thus define:
\begin{equation}
  \pi_{\theta}^{r_j}(\varpi_{\mathrm{s},j}|Y)\coloneqq
  \frac{\pi_{\theta}(\varpi_{\mathrm{s},j}|Y)^{r_j}}%
       {\int\pi_{\theta}(\varpi_{\mathrm{s},j}|Y)^{r_j}\:\dif\varpi_{\mathrm{s},j}},\qquad r_j>0.
\end{equation}
The $r_j$ parameter can be chosen separately for each tone; %
with $R=(r_1,\dotsc,r_m)$, we further define:
\begin{equation}\label{eq:liftcomb}
  \pi_{\theta}^{R}(\varpi_{\mathrm{s}}|Y)\coloneqq
  \pi_{\theta}^{r_1}(\varpi_{\mathrm{s},1}|Y)\cdot\pi_{\theta}^{r_2}(\varpi_{\mathrm{s},2}|Y,\varpi_{\mathrm{s},1})\cdots\pi_{\theta}^{r_m}(\varpi_{\mathrm{s},m}|Y,\varpi_{\mathrm{s},1},\dotsc,\varpi_{\mathrm{s},m-1}).
\end{equation}
We set $S=3^m$ and let $R_1,\dots,R_S$ be all the combinations of $m$ elements
out of the set $\{1,0.1,0.01\}$, which turn out to be reasonable magnitudes.
Additionally, we do not only accept the bias which causes an underestimation of
the policy gradient, but we also artificially scale it down by a factor of $10$.
Since the policy gradient and the backpropagation gradient model different
parameters, this is conceptually not a problem; however, if the gradient is
scaled \emph{too} much, the interdependencies of the values in the network
can then cause instabilities in the output of the stochastic parameters.
Our modified gradient estimator is:
\begin{equation}
\hat{g}_{\theta,Y}=\frac{1}{S}\sum_{i=1}^S\biggl(\frac{1}{10}\nabla_\theta\log\pi_{\theta}^{R_i}(\varpi_{\mathrm{s}}^i|Y)\cdot \bigl(L(\varpi_{\mathrm{s}}^i,\theta,D,Y)-\hat{C}(\theta,D,Y)\bigl)+\nabla_\theta L(\varpi_{\mathrm{s}}^i,\theta,D,Y)\biggr),
\end{equation}
where each tone parameter set $\varpi_{\mathrm{s},j}^i$ is sampled according
to the value of $r_j$ inside $R_i$.  This includes the sparsity parameter
$u_j$, but the inharmonicity $b_j$ is
exempt from this modification altogether and always sampled according to
$\pi_{\theta}$.

Since the way of sampling also affects
the empirical baseline $\hat{C}(\theta,D,Y)$, it is no longer an estimator
for $C(\theta,D,Y)$ but simply the mean loss among the samples.
We observe this to be a good stabilizer for the gradient.
Via our choice of $R_1,\dotsc,R_S$ we make sure that
one sample is drawn with $(r_1,\dotsc,r_m)=(1,\dotsc,1)$
from the original distribution $\pi_{\theta}$.
Also, even with increased exploration in some of the tones,
there is always a combination with $r_j=1$ for the other tones.
Therefore, rather than destabilizing all tones at the same time,
part of the exploration is only performed selectively on specific tones
with the other ones still sampled via $\pi_{\theta}$.

We also need to train the dictionary $D$; since the representation of
$D$ is not conditional but simply a variable of size
$N_{\mathrm{har}}\times N_{\mathrm{ins}}$, exploration is not desired.
In fact, even though the policy gradient on the dictionary
exists for all tones but the first
(via the dependency of $y_j$ on $D$), we ignore it to increase training
stability.
However,
the variables $\varpi_{\mathrm{s}}^i$ were sampled according to $\pi_\theta^{R_i}$
rather than $\pi_\theta$, so to get a stable estimate for $D$,
we \enquote{undo} this modification by multiplying with the ratio between
the probability densities.  We estimate the gradient in $D$ via:
\begin{equation}
  \hat{g}_{D,Y}=\frac{\sum_{i=1}^S\rho_i\,\nabla_D L(\varpi_{\mathrm{s}}^i,\theta,D,Y)}%
   {\sum_{i=1}^S\rho_i},\qquad
   \rho_i=\frac{\pi_\theta(\varpi_{\mathrm{s}}|Y)}{\pi_\theta^{R_i}(\varpi_{\mathrm{s}}|Y)},
\end{equation}
where $\pi_\theta^{R_i}$ is defined according to \eqref{eq:liftcomb}.
This method is called \emph{weighted importance sampling}
\cite[cf.][]{Sutton20}.

\subsection{Network Architecture}
\label{sec:architecture}

\begin{figure}
\small\centering
\tikzsetnextfilename{Figure_\the\numexpr(\thefigure+1)\relax}
\tikzset{
module/.style={circle, draw=black!50!white, line width=2pt},
dict/.style={rectangle, draw=black!50!white, line width=2pt},
loss/.style={rectangle, draw=black!50!white, line width=2pt},
output/.style={rectangle, draw, minimum height=1.6em, inner ysep=0pt},
var/.style={rectangle},
to_bg/.style={draw=white, line width=3pt},
to/.style={-stealth},
return/.style={-stealth, draw=black!30!white, line width=3pt},
lineto/.style={},
sample/.style={-stealth,decorate,decoration={snake}},
sample_bg/.style={draw=white, line width=3pt,decorate,decoration={snake}},
indexing/.style={dashed},
}
\begin{tikzpicture}
 \node[output] (nu_eta_params)
     {logits};
 \node[var, right=3 of nu_eta_params] (nu_eta)
     {$\nu_j,\eta_j$};
 \node[output, below=1.5 of nu_eta_params.west, anchor=west] (b_out) {};
 \node[var, right=1pt of b_out] (b_index)
     {$[\nu_j,\eta_j]$};
 \node[var, right=0.5 of b_index] (b_params)
     {$\alpha^\Gamma_j,\beta^\Gamma_j$};
 \node[var, right=3 of b_params] (b)
     {$b_j$};
 \node[output, below=of b_out.west, anchor=west] (a_out) {};
 \node[var, right=1pt of a_out] (a_index)
     {$[\nu_j,\eta_j]$};
 \node[var, right=0.5 of a_index] (a)
     {$a_j$};
 \node[output, below=of a_out.west, anchor=west] (sigma_out) {};
 \node[var, right=1pt of sigma_out] (sigma_index)
     {$[\nu_j,\eta_j]$};
 \node[var, right=0.5 of sigma_index] (sigma)
     {$\sigma_j$};
 \node[output, below=of sigma_out.west, anchor=west] (tilde_nu_out) {};
 \node[var, right=1pt of tilde_nu_out] (tilde_nu_index)
     {$[\nu_j,\eta_j]$};
 \node[var, right=0.5 of tilde_nu_index] (tilde_nu)
     {$\tilde\nu_j$};
 \node[var, right=0.6 of tilde_nu] (f)
     {$(f_{j,h})_{h=1,\dotsc,N_{\mathrm{har}}}$};
 \node[output, below=of tilde_nu_out.west, anchor=west] (v_out) {};
 \node[var, right=1pt of v_out] (v_index)
     {$[\cdot,\eta_j]$};
 \node[var, below=of tilde_nu.west, anchor=west] (v)
     {$v_j$};
 \node[var, right=0.6 of v] (c)
     {$(c_{j,h})_{h=1,\dotsc,N_{\mathrm{har}}}$};
 \node[var, right=of c] (varphi)
     {$(\varphi_{j,h})_{h=1,\dotsc,N_{\mathrm{har}}}$};
 \node[output, below=of v_out.west, anchor=west] (u_params) {};
 \node[var, right=1pt of u_params] (u_index)
     {$[\nu_j,\eta_j]$};
 \node[var, right=3 of u_index] (u)
     {$u_j$};

 \node[var, right=0.8 of varphi] (y) {$y_j$};
 \node[var, below=0.8 of varphi] (y_dir) {$y^{\mathrm{dir}}_j$};

 \node[var, dict, right=1.6 of nu_eta] (dict_entries) {$(D[h,\eta_j])_{h=1,\dotsc,N_{\mathrm{har}}}$};

 \node[var, below left=of u] (update_y_spr_l) {$y^{\mathrm{spr}}$};
 \node[var, right=0pt of update_y_spr_l] (update_y_spr_r)
     {$\leftarrow y^{\mathrm{spr}} + u_j y_j$};
 \node[var, below=0.8 of update_y_spr_l.east, anchor=east] (update_mix_l) {$y$};
 \node[var, right=0pt of update_mix_l] (update_mix_r) {$\leftarrow y + y_j$};
 \node[var, below=1.4em of update_mix_l.east, anchor=east] (update_mix_dir_l) {$y^{\mathrm{dir}}$};
 \node[var, right=0pt of update_mix_dir_l] (update_mix_dir_r)
     {$\leftarrow y^{\mathrm{dir}} + y^{\mathrm{dir}}_j$};
 \node[var, below=0.8 of update_mix_dir_l.east, anchor=east] (update_j_l) {$j$};
 \node[var, right=0pt of update_j_l] (update_j_r) {$\leftarrow j + 1$};

 \path[draw=black] (nu_eta_params.north west) -- (u_params.south west)
     coordinate[midway] (unet_outputs);
 \node[module, left=0.25 of unet_outputs] (unet) {U-Net};
 \node[var, left=0.25 of unet] (unet_inputs)
     {\shortstack{
      $Y-y$\\
      $Y-y^{\mathrm{dir}}$\\
      $|Y-y|$\\
      $|Y-y^{\mathrm{dir}}|$\\
      $y_1$\\
      $y^{\mathrm{dir}}_1$\\
      \vdots\\
      $y_{j-1}$\\
      $y^{\mathrm{dir}}_{j-1}$}};

 \path (b.south west) ++(-0.2, -0.2) coordinate (b_to_f_aux1);
 \path (f.north west) ++(0.6, 0) coordinate (b_to_f_final);
 \path (nu_eta.south west) ++(0, -0.3) coordinate (nu_eta_to_f_aux1);
 \path (f.north west) ++(0.5, 0) coordinate (nu_eta_to_f_final);
 \draw[to] (b) -- (b_to_f_aux1) -| (b_to_f_final);
 \draw[to] (nu_eta) -- (nu_eta_to_f_aux1) -| (nu_eta_to_f_final);
 \path (a.east) ++(0.8,-0.2) coordinate (a_to_y_dir_aux1);
 \coordinate (a_to_y_dir_aux2) at (a_to_y_dir_aux1 -| c.east);
 \path (a_to_y_dir_aux2) ++(0.6,0) coordinate (a_to_y_dir_aux3);
 \path (y_dir) ++(-0.8,0.3) coordinate (a_to_y_dir_aux4);
 \path (sigma.east) ++(0.8,-0.2) coordinate (sigma_to_y_dir_aux1);
 \coordinate (sigma_to_y_dir_aux2) at (sigma_to_y_dir_aux1 -| c.east);
 \path (sigma_to_y_dir_aux2) ++(0.5,0) coordinate (sigma_to_y_dir_aux3);
 \path (y_dir) ++(-0.8,0.2) coordinate (sigma_to_y_dir_aux4);
 \path (f.east) ++(0.2,-0.2) coordinate (f_to_y_dir_aux1);
 \coordinate (f_to_y_dir_aux2) at (f_to_y_dir_aux1 -| c.east);
 \path (f_to_y_dir_aux2) ++(0.4,0) coordinate (f_to_y_dir_aux3);
 \path (y_dir) ++(-0.8,0.1) coordinate (f_to_y_dir_aux4);
 \path (c.east) ++(0.2,-0.2) coordinate (c_to_y_dir_aux1);
 \path (c.east) ++(0.3,-0.2) coordinate (c_to_y_dir_aux2);
 \draw[to_bg] (sigma_to_y_dir_aux1) -- (sigma_to_y_dir_aux3);
 \draw[to] (sigma) -- (sigma_to_y_dir_aux1) -- (sigma_to_y_dir_aux3) |-
     (sigma_to_y_dir_aux4) -- (y_dir);
 \draw[to] (c) -- (c_to_y_dir_aux1) -- (c_to_y_dir_aux2) |- (y_dir);
 \draw[to] (f) -- (f_to_y_dir_aux1) -- (f_to_y_dir_aux3) |- (f_to_y_dir_aux4) --
     (y_dir);
 \coordinate (a_to_y_aux1_1) at (a -| f.east);
 \path (a_to_y_aux1_1) ++(1.2,0) coordinate (a_to_y_aux1);
 \path (y.north west) ++(-0.4,0.4) coordinate (a_to_y_aux2);
 \draw[to_bg] (a) -| (a_to_y_aux1);
 \draw[to] (a) -- (a_to_y_aux1) |- (a_to_y_aux2) -- (y);
 \draw[to] (dict_entries) -| (y);
 \path (f.east) ++(1,0) coordinate (f_to_y_aux1);
 \path (y.north west) ++(-0.4,0.2) coordinate (f_to_y_aux2);
 \draw[to_bg] (f) -- (f_to_y_aux1);
 \draw[to] (f) -- (f_to_y_aux1) |- (f_to_y_aux2) -- (y);
 \coordinate (sigma_to_y_aux1_1) at (sigma -| f.east);
 \path (sigma_to_y_aux1_1) ++(1.1,0) coordinate (sigma_to_y_aux1);
 \path (y.north west) ++(-0.4,0.3) coordinate (sigma_to_y_aux2);
 \draw[to_bg] (sigma) -- (sigma_to_y_aux1);
 \draw[to] (sigma) -- (sigma_to_y_aux1) |- (sigma_to_y_aux2) -- (y);
 \draw[to] (varphi) -- (y);
 \draw[sample] (nu_eta_params) -- node[above] {\strut Categorical} (nu_eta);
 \draw[lineto] (b_index) -- (b_params);
 \draw[sample_bg] (b_params) -- (b);
 \draw[sample] (b_params) -- node[above]
     {$\Gamma(\alpha^\Gamma_j, \beta^\Gamma_j)$} (b);
 \draw[lineto] (a_index) -- (a);
 \draw[lineto] (sigma_index) -- (sigma);
 \draw[lineto] (tilde_nu_index) -- (tilde_nu);
 \draw[lineto] (v_index) -- (v);
 \draw[sample] (u_index) -- node[above]
     {\strut Bernoulli} (u);
 \draw[to] (tilde_nu) -- (f);
 \draw[to] (v) -- (c);
 \draw[to_bg] (c) -- (varphi);
 \draw[to] (c) -- (varphi);
 \draw[to, indexing] (nu_eta) -- (dict_entries);
 \path (nu_eta.south west) ++(-1,-0.2) coordinate (nu_eta_to_indices_aux1);
 \draw[to, indexing] (nu_eta) -- (nu_eta_to_indices_aux1) -| (b_index);
 \path (y) ++(-0.2,-1) coordinate (y_to_update_y_spr_r_aux1);
 \coordinate (y_to_update_y_spr_r_aux2) at
     (y_to_update_y_spr_r_aux1 |- update_y_spr_r);
 \coordinate (y_to_update_mix_r_aux1) at (y |- update_mix_r);
 \path (update_y_spr_r.north) ++(16pt,0) coordinate (u_to_update_y_spr_r_final);
 \draw[to] (u) -| (u_to_update_y_spr_r_final);
 \draw[to] (y_dir) |- (update_mix_dir_r);
 \draw[to_bg] (y_to_update_y_spr_r_aux2) -- (update_y_spr_r);
 \draw[to] (y) -- (y_to_update_y_spr_r_aux1) -- (y_to_update_y_spr_r_aux2) --
     (update_y_spr_r);
 \draw[to_bg] (y_to_update_mix_r_aux1) -- (update_mix_r);
 \draw[to] (y) -- (y_to_update_mix_r_aux1) -- (update_mix_r);
 \path (update_mix_dir_l.north west) ++(-0.5,0) coordinate (return_start);
 \path (unet_inputs.south) ++(0,-0.5) coordinate (return_final);
 \draw[return] (return_start) -| (return_final);
 \draw[to] (unet_inputs) -- (unet);
 \draw[to] (unet) -- (unet_outputs);
\end{tikzpicture}
\caption{Sampling architecture}
\label{fig:arch}
\end{figure}
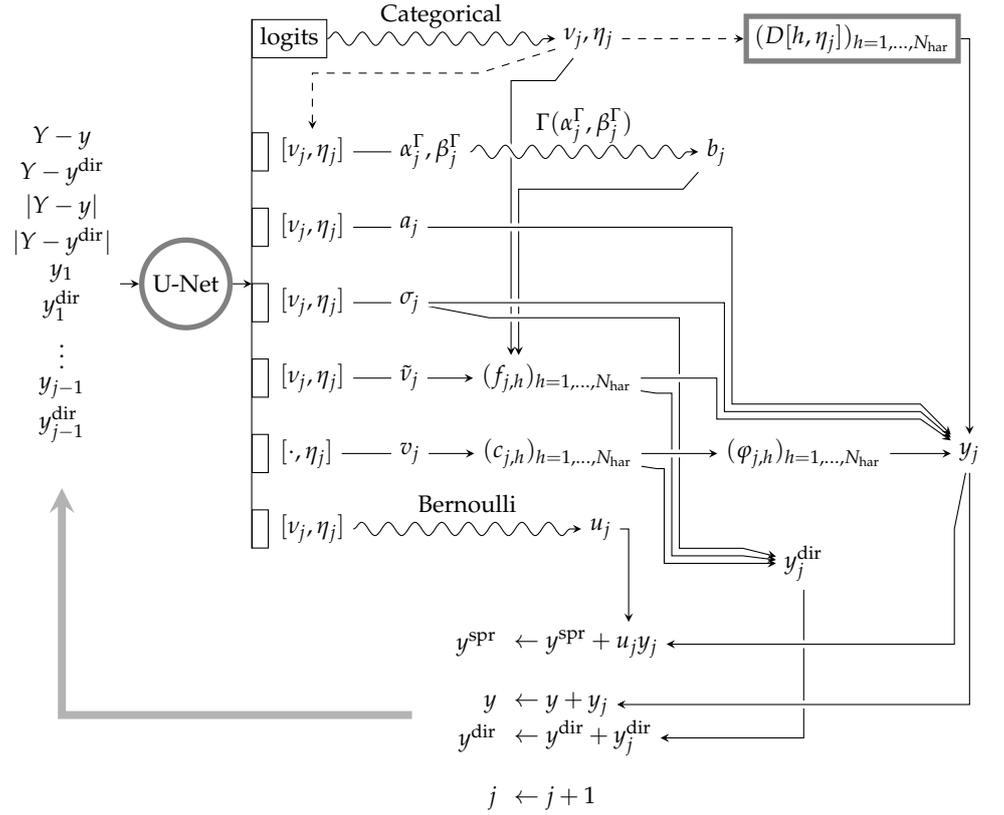

We use a U-Net architecture with 7 downsampling/upsampling steps;
strides of $4,4,4,4,4,3,2$; and $80,160,\dotsc,560$ one-dimensional filters
with a size of $5$.  Finally, we add two more convolutional layers
with $80$ filters of sizes $3,1$, respectively, before the linear output layer.
All of the hidden layers have ReLU activation.  The first hidden layer
is a \emph{CoordConv} layer \cite{Liu18},
which means that it is supplied with a linear range
from $0.01$ to $0$ as an additional input channel.
These design parameters were obtained via manual experimentation.

For the first tone, the network is supplied with the input spectrum $Y$ and the
absolute value spectrum $\sabs{Y}$.  From this, the spectra $y_1$ and
$y_1^{\mathrm{dir}}$ are computed.  For each following tone $j=2,\dotsc,m$,
the network receives the residuals $Y-y_1-\ldots-y_{j-1}$ and
$Y-y_1^{\mathrm{dir}}-\ldots-y_{j-1}^{\mathrm{dir}}$ along with
their absolute values as well as the computed tone spectra
$y_1,\dotsc,y_{j-1}$ and $y_1^{\mathrm{dir}},\dotsc,y_{j-1}^{\mathrm{dir}}$.
From these values, it then yields the spectra $y_j$ and $y_j^{\mathrm{dir}}$.
For those input components of the network that do not yet receive a
spectrum for a particular tone, a constant $0$ vector is given as input to
the network instead.  The data flow is illustrated in Figure \ref{fig:arch}.

Since the amplitudes $(a_j)$ are supposed to be non-negative, we apply
the absolute value function to the respective output components of the network.
The widths $(\sigma_j)$ are kept
positive via softplus, and they are clipped such that the value does not
get too close to $0$.  For the continuous frequency offsets
$(\tilde{\nu_j})$, a $\tanh$ function is used to keep them
inside the interval $(-5,5)$.  The positive parameters
$(\alpha_j^\Gamma,\beta_j^\Gamma)$
are obtained after applying the exponential function, and the probabilities
for the Bernoulli distribution for the sparsity parameters $(u_j)$
are mapped into the interval $(0,1)$ via a sigmoid function.
The joint categorical distribution for the
discrete frequencies $(\nu_j)$ and instrument indices $(\eta_j)$ is given
in vectorial form as non-normalized log-probabilities, so we apply the
\emph{softmax} mapping in order to obtain a valid discrete distribution.
Doing this, we have to
make sure that each instrument can only play one tone at a time
by excluding that instruments that have already been assigned a tone from
the sampling.
For each parameter output by the network, we add a trainable scaling layer.

In order to protect against potentially degenerate network output or gradients
in the case
of zeros in the input vector, a minimal amount of Gaussian noise is always
added to $Y$ prior to prediction, on a level that is negligable for any
normal audio signals.

The dictionary entries are supposed to be elements of the interval $[0,1]$.
Non-negativity is usually satisfied automatically due to $a_j\geq0$,
and for the upper bound, we add the loss
$\frac{1}{N_{\mathrm{ins}}} \sum_\eta(\log (\max_h D[h,\eta]))^2$ to the
training of the dictionary.

\subsection{Training}
\label{sec:training}

The network weights are Glorot-initialized and the biases are initially
set to $0$.  The initial values for the instruments in the dictionary
are exponentially decaying sequences along the harmonics:
\begin{equation}
  D_0[h,\eta]
 =\Bigl(\frac{0.5}{\eta}\Bigr)^{h-1}.
\end{equation}

We partition the spectra $Z[k,\cdot]$, $k=1,\dotsc,n_{\mathrm{len}}$,
into random batches of size $6$.  We then train on each batch with the
AdaMax algorithm \cite{Kingma14}.
For each epoch, new random batches are assigned.
For the dictionary, we also use AdaMax, but with a reduced learning rate
of $\num{e-4}$.  Also, analogously to \cite{Schulze21}, for the denominator
in AdaMax, we consider the maximum over all the harmonics of the
particular instrument when training $D$.  The entire procedure is
outlined in Algorithm \ref{algo:adamax}.

\begin{algoenv}[hbtp]
  \centering
\begin{minipage}[t]{\columnwidth}
  \caption{Training scheme for the network and the dictionary,
    based on AdaMax \cite{Kingma14}.
  Upper bound regularization of $D$ and batch summation (see Sections \ref{sec:architecture}, \ref{sec:training}) are not explicitly stated.}\label{algo:adamax}
\centering%
\begin{varwidth}{1\textwidth}
  \textbf{Input:}\ $Z$, $\theta$, $D$\\
  \textbf{Parameters:}\ $T\in\N$, $\kappa_\theta>0$, $\kappa_D>0$, $\beta_1\in(0,1)$, $\beta_2\in(0,1)$, $\eps>0$
  \begin{algorithmic}
    \State $\gamma_{\theta,1}\leftarrow 0$
    \State $\gamma_{\theta,2}\leftarrow 0$
    \State $\gamma_{D,1}\leftarrow 0$
    \State $\gamma_{D,2}\leftarrow 0$
    \For{$\tau=1,\dotsc,T$}
      \State choose $Y$ out of $\{Z[k,\cdot]:k=1,\dotsc,n_{\mathrm{len}}\}$
      \State $\gamma_{\theta,1}\leftarrow\beta_1\,\gamma_{\theta,1}+(1-\beta_1)\,\hat{g}_{\theta,Y}$
      \State $\gamma_{\theta,2}\leftarrow\max(\beta_2\,\gamma_{\theta,2},\sabs{\hat{g}_{\theta,Y}})$
      \State $\theta\leftarrow\theta-\frac{\kappa_\theta}{1-\beta_1^\tau}\cdot\frac{\gamma_{\theta,1}}{\gamma_{\theta,2}+\eps}$
      \State $\gamma_{D,1}\leftarrow\beta_1\,\gamma_{D,1}+(1-\beta_1)\,\hat{g}_{D,Y}$
      \State $\gamma_{D,2}\leftarrow\max(\beta_2\,\gamma_{D,2},\max_h\sabs{\hat{g}_{D,Y}[h,\cdot]})$
      \State $D\leftarrow D-\frac{\kappa_D}{1-\beta_1^\tau}\cdot\frac{\gamma_{D,1}}{\gamma_{D,2}+\eps}$
    \EndFor
  \end{algorithmic}
  \textbf{Output:}\ $\theta,D$
\end{varwidth}
\end{minipage}

\vspace{2ex}
Typical choice:
$N=\num{70000}$, $\kappa_\theta=\num{e-3}$, $\kappa_D=\num{e-4}$, $\beta_1=0.9$, $\beta_2=0.999$, $\eps=\num{e-7}$
\end{algoenv}

\subsection{Resynthesis}

After training,
we apply the network once again on all the time frames $Z[k,\cdot]$,
$k=1,\dotsc,m$.  To prevent randomness in the output,
rather than sampling according to $\pi_\theta(\varpi_{\mathrm{s}}|Y)$, we 
use the mode of $\pi_\theta(\nu_j,\eta_j|Y)$ and then, with $\nu_j,\eta_j$
fixed, the modes of
$\pi_\theta(b_j|Y,\nu_j,\eta_j)$ and of $\pi_\theta(u_j|Y,\nu_j,\eta_j)$.

We project the thereby obtained
time-frequency coefficients $z^{\mathrm{dir}}_j[k,l]$ back
into real-valued time-domain signals for each instrument
via \cite[cf.][]{Groechenig01,Doerfler02}:
\begin{equation}
x_j^{\mathrm{syn}}(t)=\sum_{k,l}z_j^{\mathrm{dir}}[k,l]\,
\tilde{w}(t-\alpha k)\,e^{i2\pi\beta l(t-\alpha k)},
\end{equation}
where
\begin{equation}
\tilde{w}(t)=
\frac{\beta\,w(t)}{\sum_k\sabs{w(t-\alpha k)}^2}
\end{equation}
is the \emph{synthesis window}.  If the support of $w$ is cut to
$[-1/(2\beta),1/(2\beta)]$, then it is also the
\emph{Gabor canonical dual window} of $w$,
which is the miminum-$L_2$-norm window to invert the transformation
from $X$ to $Z$ according to \eqref{eq:sample}.

\section{Experimental Results and Discussion}

We compare our algorithm against two other blind source separation algorithms.
We selected them for their ability to identify the sound of musical
instruments at arbitrary pitch on a continuous frequency axis.
\begin{enumerate}
\item The algorithm from a previous publication
  of some of the authors \cite{Schulze21} assumes
  an identical tone model, but instead of a trained neural network,
  it uses a hand-crafted sparse pursuit algorithm for identification,
  and it operates on a specially computed log-frequency spectrogram.
  While the data model can represent inharmonicity, it is not fully
  incorporated into the pursuit algorithm.  Also, information is lost
  in the creation of the spectrogram.  Since the algorithm operates completely
  in the real domain, it does not consider phase information, which
  can lead to problems in the presence of beats.
  The conceptual advantage of the method is that it only requires rather
  few hyperparameters and their choice is not critical.
\item The algorithm by Duan et al.\ \cite{Duan08} detects and clusters peaks
  in a linear-frequency STFT spectrogram via a probabilistic model.
  Its main advantage over other methods is that it can extract instrumental
  music out of a mixture with signals that cannot be represented.
  However, this comes at the cost of having to tune the parameters
  for the clustering algorithm specifically for every sample.
\end{enumerate}

While hyperparameter choice is more important for our new algorithm
than for the first algorithm from this list, we aim to maintain our notion
of blind separation by keeping our choice constant for
all the samples that we consider, while for any comparison to the
second algorithm, it should be kept in mind that the hyperparameters for this
method are hand-optimized with the values taken from
\cite{Duan08} and \cite{Schulze21}.  When comparing to the algorithm
from \cite{Schulze21}, we always consider the result with spectral masking
applied.

For reconstruction, unless otherwise stated, we use the values from
\eqref{eq:default}.  However, in training, we aim to increase the
amount of training data available to the network (\emph{data augmentation})
by using a modified time constant $\tilde{\alpha}=\alpha/4$.  While
the newly added spectra are completely redundant and do not add any information
to the original ones, this connection is not built into the neural network
and therefore the redundant training data can help it learn details that
it would not learn from a representation without this redundancy.

For all the samples with $2$ instruments, we train for $\num{100000}$ iterations
(independently of the size of one epoch),
but we always use the result after $\num{70000}$ iterations (\emph{early stopping}),
since it appears that this usually gives better results.  We conjecture that
the partially trained network itself provides a good regularization to which
separations are realistic, while simply minimizing the loss itself can
lead to degenerate results.  Regularization via network architecture in
unsupervised learning has been prominently pioneered via the
deep image prior approach \cite{Ulyanov18}.

For assessing separation quality, we use the SDR (signal-to-distortion ratio),
which measures the overall similarity between the original and the
resynthesized signal, the SIR (signal-to-interference ratio),
which gives the interference from the other instrument tracks in the
considered signal, and the SAR (signal-to-artifacts ratio), which
disregards interference and compares the given signal to all the original ones
\cite{Vincent06}.
These are well-established figures in blind separation from the
\emph{BSS Eval} software package \cite{Fevotte05}.\footnote{%
  We follow the definitions compatible with version 2 of BSS Eval,
  which, unlike version 3, does not permit shifts in the signal.}
For all of them, a higher number corresponds to better quality.

When training non-trivial neural networks, it is always,
to varying extent, a matter
of chance if the optimization process will converge to a good value.
We therefore train an \emph{ensemble} of neural networks by running the process
with $6$ different random seeds (affecting the network initialization,
the batch partitioning, and the random sampling of the stochastic
parameters)\footnote{%
  These seeds are distinct from those that we used for
  validation and hyperparameter selection during the design process
  of the algorithm.}
and choose the best one in terms of mean SDR over the instruments.
In a realistic blind scenario without ground truth, this figure would
not be available, but we deem it acceptable to let the user choose
(for instance, by listening comparison) between
a small number of different output results.  Due to the aforementioned
regularization by architecture, the value of the loss function is not a
reliable indicator of separation quality, but it could be integrated
into an end-user interface.

Similarly, the results taken from \cite{Schulze21} are always the best-case
training outcomes out of 10 runs.  By contrast, the algorithm from
\cite{Duan08} does not rely on randomness but instead on the hand-optimized
hyperparameters.

\subsection{Mozart's Duo for Two Instruments}

Like in \cite{Schulze21}, we use the 8th piece from the 12 Basset Horn Duos by
Wolfgang A.\ Mozart (K.~487) in an arrangement by Alberto Gomez Gomez
for two recorders\footnote{%
  \url{https://imslp.org/wiki/12_Horn_Duos,_K.487/496a_(Mozart,_Wolfgang_Amadeus)}}
as an example piece.  In one sample, it is played with recorder and violin,
and in the second sample, it is played with clarinet and piano.
All the instruments are acoustic.

In Table \ref{tab:mozart}, we compare the performance of our algorithm
to that of the other two on both samples.  The sample with recorder and violin
is comparatively \enquote{easy.}  While our proposed algorithm universally
gives the best SIR values (indicating low interference between the
instrument tracks), the algorithm from \cite{Schulze21}
outperforms it for the recorder track in terms of
SDR and SAR.
Possible explanations include a more effective optimization process of the
respective objective function,
but it could also be a result of the different data representation used
in that method (namely the special log-frequency spectrogram)
or a regularizing effect of the sparse pursuit algorithm.
In preliminary experiments, we found that in this particular sample,
a lower value for $\mu_1$ can increase separation quality
(putting more emphasis on the correctness of the direct prediction
rather than the dictionary-based prediction), but requiring
the user to guess the difficulty beforehand violates our notion of
blind separation.

The first few seconds of the separation result are displayed in spectrogram
form in Figure~\ref{fig:mozart}.  Overall, the direct predictions are very
accurate, but like it was observed in \cite{Schulze21}, the last tone visible in
Figure~\ref{fig:mozart}b jumps up a fifth, which it does not do in
the ground truth.  This is because the recorder tone is actually one
octave above the violin tone, and the overlap is virtually perfect.
In such cases, since the dictionary model is never fully accurate, it
may happen that an incorrect solution actually yields a lower loss.

By comparison, the sample with clarinet and piano is rather \enquote{hard,}
and our algorithm clearly shows superior performance, especially in
the separation quality of the piano track.  An interesting observation
in Figure~\ref{fig:mozcl}a
is that while the separation performance first reaches a plateau around
\SI{10}{dB} (in the mean), it then declines to around \SI{8}{dB}.
However, as shown in Figure~\ref{fig:mozcl}b, the values of both
$d_{2,\delta}^{q,\mathrm{abs}}(Y,y^{\mathrm{spr}})$ and
$d_{2,\delta}^{q,\mathrm{rad}}(Y,y^{\mathrm{dir}})$ decrease, indicating this
is not an optimization failure.  At the same time, the value of the
regularization loss (which, with $\mu_3=1$, has a much lower weight than
the others) increases.  Therefore, this particular sample could potentially
benefit from more reguralization.

In Figure~\ref{fig:mozarttrain}, the learning curves for both samples
over all the respective runs are displayed.  It is obvious that
the results for the sample with recorder and violin
(Figure~\ref{fig:mozarttrain}a) are more consistent than those for the
sample with clarinet and piano (Figure~\ref{fig:mozarttrain}b).
Also, we can see that in the former sample, separation quality generally
deteriorates after too many iterations while in the latter, some runs
only achieve peak performance near the end of the training.  Thus,
different samples can benefit from taking the results at different
points in the training process.
Some curves in Figure~\ref{fig:mozarttrain}b terminate early, which
was due to numerical failures in the training process.

\begin{specialtable}[ht]
  \centering
  \caption{Comparison of the separation algorithms on
    the samples based on the piece by Mozart. Best numbers are marked.}
    \label{tab:mozart}
    \tabcolsep=4.5pt
    \renewcommand{\arraystretch}{1.1}
    \heavyrulewidth=.5pt
    \lightrulewidth=.4pt
    \cmidrulewidth=.3pt
    \aboverulesep=0.0ex
    \belowrulesep=0.0ex

  \newcolumntype Y{S[
      table-format=2.1,
      table-auto-round,
      table-text-alignment=center,
      table-number-alignment=center,
      table-space-text-post={*}]}
    \begin{tabular}{ccYYY}
    \toprule
    {\bfseries Method} & {\bfseries Instrument} & {\bfseries SDR} & {\bfseries SIR} & {\bfseries SAR} \\
    \midrule
    \multirow{4}{*}{Ours} & Recorder & 13.1 & 34.8* & 13.2 \\
    & Violin & 13.4* & 34.2* & 13.5* \\
    \cmidrule{2-5}
    & Clarinet & 12.4* & 28.0* & 12.6* \\
    & Piano & 8.1* & 42.2* & 8.1*\\
    \cmidrule{1-5}
    \multirow{4}{*}{\cite{Schulze21}} & Recorder & 15.13270565* & 32.38531432 & 15.21774855* \\
    & Violin & 11.86338139 & 23.7800286 & 12.1702504 \\
    \cmidrule{2-5}
    & Clarinet & 4.05190929 & 24.30443735 & 4.10916925 \\
    & Piano & 2.11202768 & 9.28523716 & 3.52042693 \\
    \cmidrule{1-5}
    \multirow{4}{*}{\cite{Duan08}} & Recorder & 10.6024 & 21.4279 & 11.0084 \\
    & Violin & 5.8343 & 18.4353 & 6.1416\\
    \cmidrule{2-5}
    & Clarinet & 6.7415 & 21.3302 & 6.9270 \\
    & Piano & 5.4587 & 16.3777 & 5.9240 \\
    \bottomrule
  \end{tabular}
\end{specialtable}

\end{paracol}
\nointerlineskip
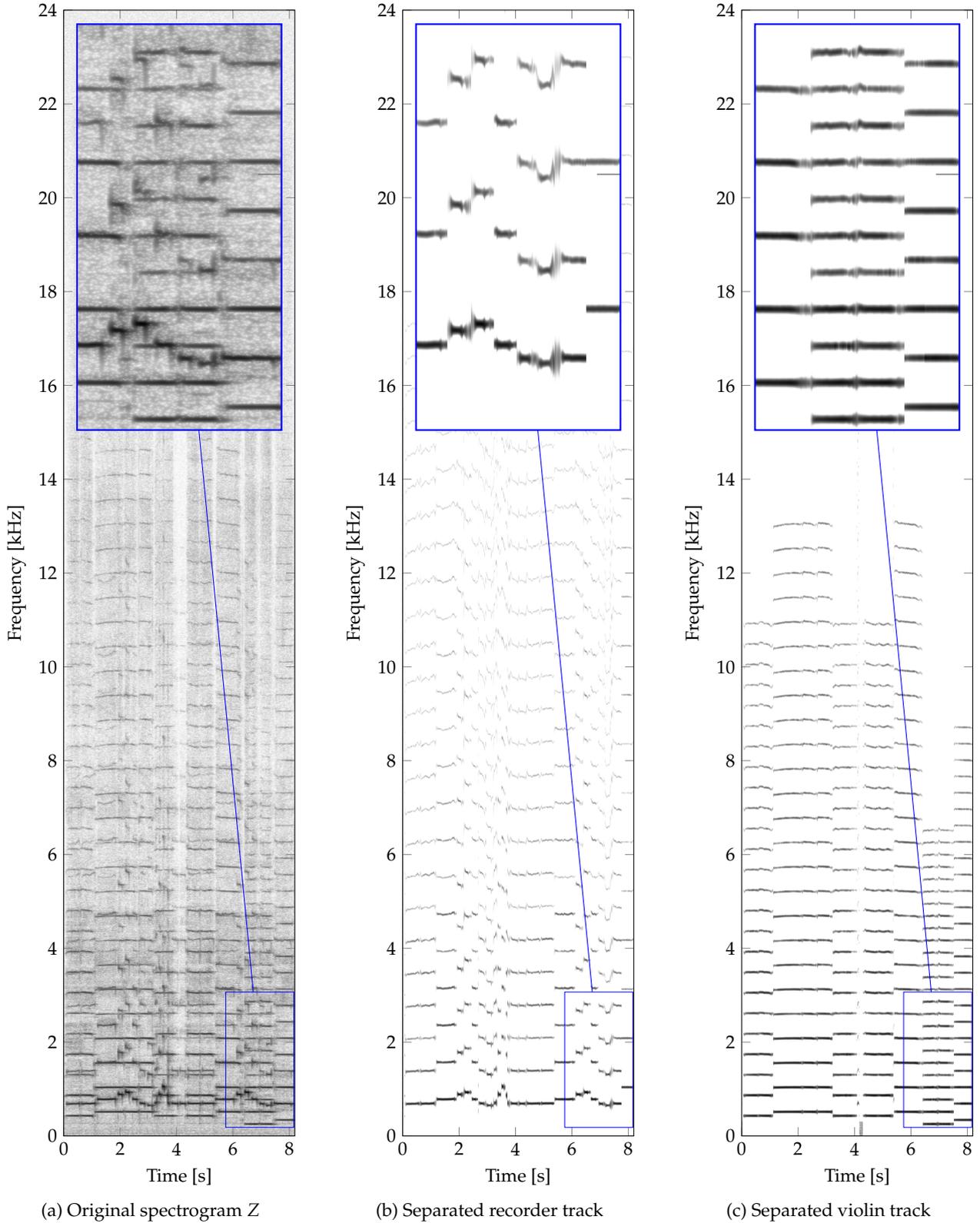
\begin{figure}
  \widefigure
  \small\centering
  \tikzsetnextfilename{Figure_\the\numexpr(\thefigure+1)\relax}
  \begin{tikzpicture}[spy using outlines={rectangle, connect spies}]
    \begin{groupplot}[group style={group size=3 by 1,horizontal sep=0.1\textwidth,vertical sep=3.75em},
        width=0.3\textwidth,title style={yshift=-1ex},
        xlabel={Time [$\si{s}$]},
        ylabel={Frequency [$\si{kHz}$]},
        height=21cm,enlargelimits=false,axis on top]
      \nextgroupplot
      \addplot graphics[xmin=0,xmax=8.1920,ymin=0,ymax=24] {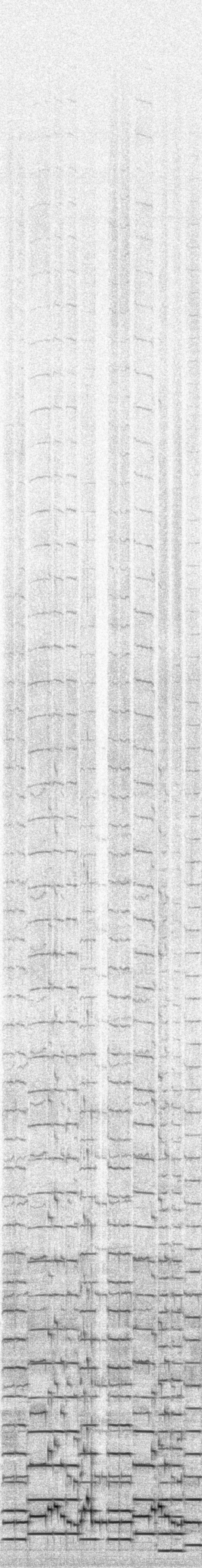};
      \coordinate (spy_on_orig) at (6.95,1.625);
      \coordinate (spy_in_orig) at (4.096,23.7);
      \spy [lens={scale=3}, width=3.5cm, height=7cm, blue] on (spy_on_orig) in node [below] at (spy_in_orig);
      \nextgroupplot
      \addplot graphics[xmin=0,xmax=8.1920,ymin=0,ymax=24] {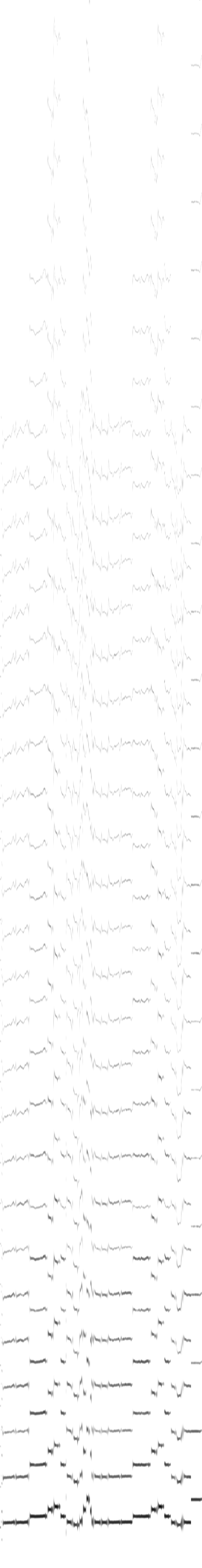};
      \coordinate (spy_on_rec) at (6.95,1.625);
      \coordinate (spy_in_rec) at (4.096,23.7);
      \spy [lens={scale=3}, width=3.5cm, height=7cm, blue] on (spy_on_rec) in node [below] at (spy_in_rec);
      \nextgroupplot
      \addplot graphics[xmin=0,xmax=8.1920,ymin=0,ymax=24] {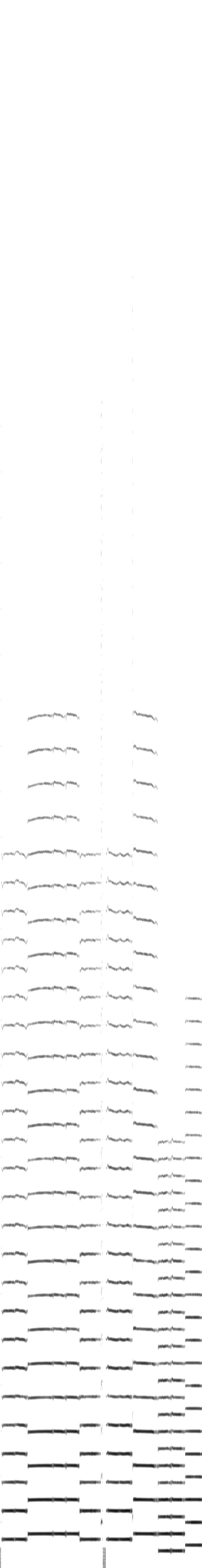};
      \coordinate (spy_on_vln) at (6.95,1.625);
      \coordinate (spy_in_vln) at (4.096,23.7);
      \spy [lens={scale=3}, width=3.5cm, height=7cm, blue] on (spy_on_vln) in node [below] at (spy_in_vln);
    \end{groupplot}
    \node[below=1mm] at (group c1r1.outer south) {(a) Original spectrogram $Z$};
    \node[below=1mm] at (group c2r1.outer south) {(b) Separated recorder track};
    \node[below=1mm] at (group c3r1.outer south) {(c) Separated violin track};
  \end{tikzpicture}
  \caption{Excerpt of the separation result for the piece by Mozart,
    played on recorder
    and violin.  Displayed are the original STFT magnitude spectrogram
    as well as the direct predictions for each instrument.
    In the highlighted section, the last tone is supposed
        to be a constant octave interval between the violin and the
        recorder, but the prediction for the recorder contains an
        erroneous jump.
    The color axes of the plots are
    normalized individually to a dynamic range of $\SI{100}{dB}$.}
  \label{fig:mozart}
\end{figure}
\begin{paracol}{2}
\makeatletter\ifthenelse{\equal{\@status}{submit}}{\linenumbers}{}\makeatother
\switchcolumn

\end{paracol}
\nointerlineskip
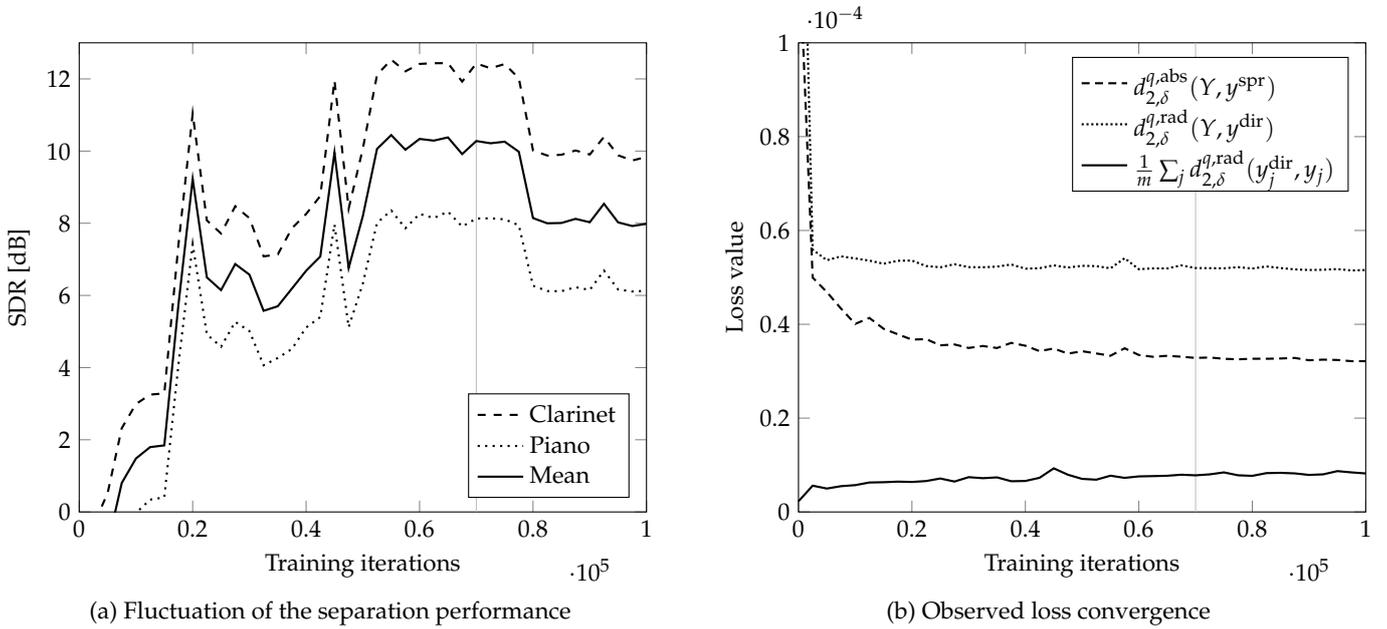
\begin{figure}[p]
  \widefigure
  \small\centering
  \tikzsetnextfilename{Figure_\the\numexpr(\thefigure+1)\relax}
  \begin{tikzpicture}
    \begin{groupplot}[group style={group size=2 by 1, horizontal sep=2cm}]
      \nextgroupplot[
            width=0.49\textwidth,enlargelimits=false,ymin=0,ymax=13,xmin=0,xmax=100000,
            legend pos=south east,legend cell align=left,
            xlabel=Training iterations,ylabel={SDR [dB]}]
        \draw[color=gray!50!white] (axis cs:70000,0) -- (axis cs:70000,13);
        \addplot[thick,dashed] table
                [x expr=\thisrowno{0}*2500,
                y expr=\thisrowno{3}] {mozart-cl/mozart-real-s12.12-measures.dat};
        \addlegendentry{Clarinet}
        \addplot[thick,dotted] table
                [x expr=\thisrowno{0}*2500,
                y expr=\thisrowno{4}] {mozart-cl/mozart-real-s12.12-measures.dat};
        \addlegendentry{Piano}
        \addplot[thick] table
                [x expr=\thisrowno{0}*2500,
                y expr=(\thisrowno{3}+\thisrowno{4})/2] {mozart-cl/mozart-real-s12.12-measures.dat};
        \addlegendentry{Mean}
      \nextgroupplot[
            width=0.49\textwidth,enlargelimits=false,xmin=0,xmax=100000,
            legend pos=north east,legend cell align=left,
            xlabel=Training iterations,ylabel={Loss value},
          ymin=0,ymax=1e-4]
        \draw[color=gray!50!white] (axis cs:70000,0) -- (axis cs:70000,1e-4);
        \addplot[thick,densely dashed] table
                [x expr=\thisrowno{0},
                y expr=\thisrowno{1}] {mozart-cl-s12.12-losses.dat};
        \addlegendentry{$d_{2,\delta}^{q,\mathrm{abs}}(Y,y^{\mathrm{spr}})$}
        \addplot[thick,densely dotted] table
                [x expr=\thisrowno{0},
                y expr=\thisrowno{2}] {mozart-cl-s12.12-losses.dat};
        \addlegendentry{$d_{2,\delta}^{q,\mathrm{rad}}(Y,y^{\mathrm{dir}})$}
        \addplot[thick] table
                [x expr=\thisrowno{0},
                y expr=\thisrowno{3}] {mozart-cl-s12.12-losses.dat};
        \addlegendentry{$\frac{1}{m}\sum_jd_{2,\delta}^{q,\mathrm{rad}}(y^{\mathrm{dir}}_j,y_j)$}
    \end{groupplot}
    \node[below=1mm] at (group c1r1.outer south) {(a) Fluctuation of the separation performance};
    \node[below=1mm] at (group c2r1.outer south) {(b) Observed loss convergence};
  \end{tikzpicture}
  \caption{Separation performance and loss values while training on the
    sample with clarinet and piano in the best-case run.  The vertical gray
    lines indicate the point at which the result was taken (\num{70000} iterations).}
  \label{fig:mozcl}
\end{figure}
\begin{paracol}{2}
\makeatletter\ifthenelse{\equal{\@status}{submit}}{\linenumbers}{}\makeatother
\switchcolumn

\end{paracol}
\nointerlineskip
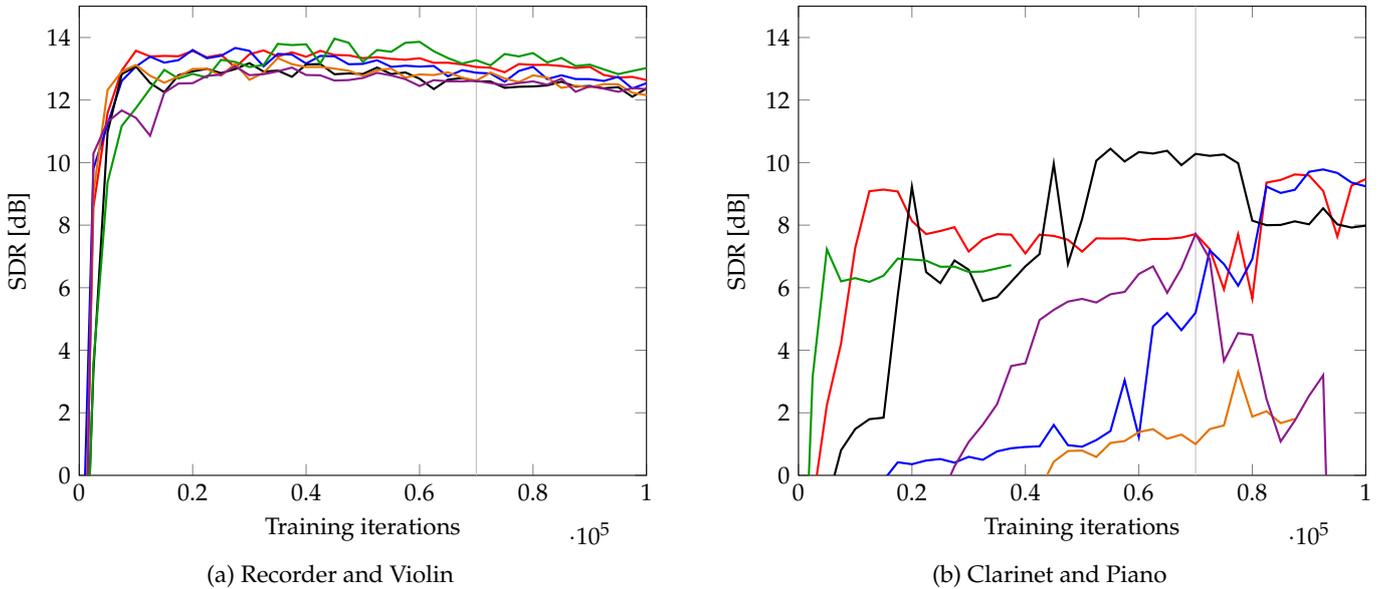
\begin{figure}[p]
  \widefigure
  \small\centering
  \tikzsetnextfilename{Figure_\the\numexpr(\thefigure+1)\relax}
  \begin{tikzpicture}
    \begin{groupplot}[group style={group size=2 by 1, horizontal sep=2cm}]
      \nextgroupplot[
            width=0.49\textwidth,enlargelimits=false,ymin=0,ymax=15,xmin=0,xmax=100000,
            legend pos=south east,legend cell align=left,
            xlabel=Training iterations,ylabel={SDR [dB]}]
        \draw[color=gray!50!white] (axis cs:70000,0) -- (axis cs:70000,15);
        \addplot[thick,red] table
                [x expr=\thisrowno{0}*2500,
                y expr=(\thisrowno{3}+\thisrowno{4})/2] {mozart/mozart-0+10abs+1+10-mod1e-4-s10.10-measures.dat};
        \addplot[thick,blue] table
                [x expr=\thisrowno{0}*2500,
                y expr=(\thisrowno{3}+\thisrowno{4})/2] {mozart/mozart-0+10abs+1+10-mod1e-4-s11.11-measures.dat};
        \addplot[thick] table
                [x expr=\thisrowno{0}*2500,
                y expr=(\thisrowno{3}+\thisrowno{4})/2] {mozart/mozart-0+10abs+1+10-mod1e-4-s12.12-measures.dat};
        \addplot[thick,green!60!black] table
                [x expr=\thisrowno{0}*2500,
                y expr=(\thisrowno{3}+\thisrowno{4})/2] {mozart/mozart-0+10abs+1+10-mod1e-4-s13.13-measures.dat};
        \addplot[thick,orange!90!black] table
                [x expr=\thisrowno{0}*2500,
                y expr=(\thisrowno{3}+\thisrowno{4})/2] {mozart/mozart-0+10abs+1+10-mod1e-4-s14.14-measures.dat};
        \addplot[thick,violet!90!white] table
                [x expr=\thisrowno{0}*2500,
                y expr=(\thisrowno{3}+\thisrowno{4})/2] {mozart/mozart-0+10abs+1+10-mod1e-4-s15.15-measures.dat};
      \nextgroupplot[
            width=0.49\textwidth,enlargelimits=false,ymin=0,ymax=15,xmin=0,xmax=100000,
            legend pos=south east,legend cell align=left,
            xlabel=Training iterations,ylabel={SDR [dB]}]
        \draw[color=gray!50!white] (axis cs:70000,0) -- (axis cs:70000,15);
        \addplot[thick,red] table
                [x expr=\thisrowno{0}*2500,
                y expr=(\thisrowno{3}+\thisrowno{4})/2] {mozart-cl/mozart-real-s10.10-measures.dat};
        \addplot[thick,blue] table
                [x expr=\thisrowno{0}*2500,
                y expr=(\thisrowno{3}+\thisrowno{4})/2] {mozart-cl/mozart-real-s11.11-measures.dat};
        \addplot[thick] table
                [x expr=\thisrowno{0}*2500,
                y expr=(\thisrowno{3}+\thisrowno{4})/2] {mozart-cl/mozart-real-s12.12-measures.dat};
        \addplot[thick,green!60!black] table
                [x expr=\thisrowno{0}*2500,
                y expr=(\thisrowno{3}+\thisrowno{4})/2] {mozart-cl/mozart-real-s13.13-measures.dat};
        \addplot[thick,orange!90!black] table
                [x expr=\thisrowno{0}*2500,
                y expr=(\thisrowno{3}+\thisrowno{4})/2] {mozart-cl/mozart-real-s14.14-measures.dat};
        \addplot[thick,violet!90!white] table
                [x expr=\thisrowno{0}*2500,
                y expr=(\thisrowno{3}+\thisrowno{4})/2] {mozart-cl/mozart-real-s15.15-measures.dat};
    \end{groupplot}
    \node[below=1mm] at (group c1r1.outer south) {(a) Recorder and Violin};
    \node[below=1mm] at (group c2r1.outer south) {(b) Clarinet and Piano};
  \end{tikzpicture}
  \caption{Mean separation performance over the instruments in the samples
    based on the piece by Mozart.  Each line represents a different
    run with specific random seeds.  The vertical gray lines
    indicate the point at which the result was taken (\num{70000} iterations).}
  \label{fig:mozarttrain}
\end{figure}
\begin{paracol}{2}
\makeatletter\ifthenelse{\equal{\@status}{submit}}{\linenumbers}{}\makeatother
\switchcolumn

One difficulty with the piano as an instrument is that it exhibits
significant inharmonicity \cite[cf.][]{Fletcher98}.  While the algorithm from
\cite{Schulze21} has been shown to correctly identify the inharmonicity
parameter on the isolated piano track, it relies on cross-correlation
\emph{without} inharmonicity for tone detection.  By contrast, the algorithm
presented here is based on a neural network, so correctly dealing with
inharmonicity at the input stage is merely a matter of training.
Since inharmonicity mostly affects higher harmonics which have low volume,
the ability to represent it in the output stage does not show up as much
in the $\ell_2$-based SDR/SAR/SIR figures.  However, due to the lifting
property of $d_{2,\delta}^{q,\mathrm{rad}}$ and $d_{2,\delta}^{q,\mathrm{abs}}$
(with $q=0.5<1$), it does influence the losses.

To illustrate the effect, we also ran our algorithm on the isolated piano
track, once with and once without the inharmonicity parameter in the model,
with $4$ distinct random seeds each.  The results for the random seeds of
$11$ are displayed in Figure \ref{fig:piano}.  While the difference
appears small on a global scale, it is consistent.

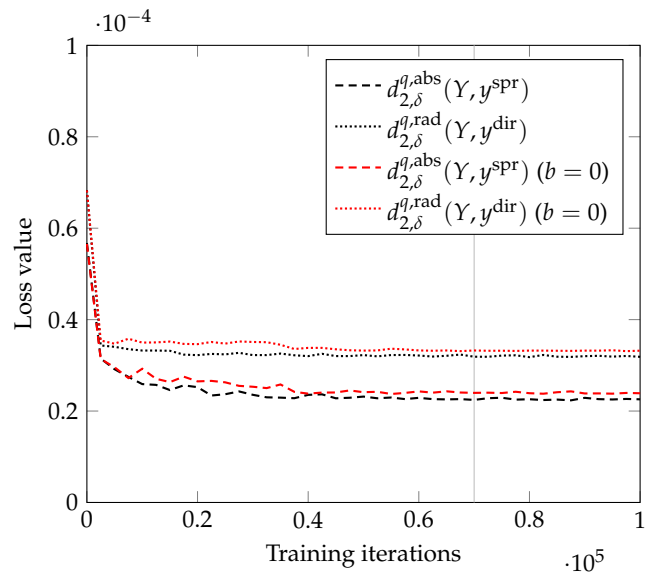
\begin{figure}[h]\centering\small
  \tikzsetnextfilename{Figure_\the\numexpr(\thefigure+1)\relax}
  \begin{tikzpicture}
    \begin{axis}[width=0.48\textwidth,enlargelimits=false,xmin=0,xmax=100000,
        legend pos=north east,legend cell align=left,
        xlabel=Training iterations,ylabel={Loss value},
        ymin=0,ymax=1e-4]
      \draw[color=gray!50!white] (axis cs:70000,0) -- (axis cs:70000,1e-4);
      \addplot[thick,densely dashed] table
              [x expr=\thisrowno{0},
                y expr=\thisrowno{1}] {mozart-cl-s11.11.dat};
      \addlegendentry{$d_{2,\delta}^{q,\mathrm{abs}}(Y,y^{\mathrm{spr}})$}
      \addplot[thick,densely dotted] table
              [x expr=\thisrowno{0},
                y expr=\thisrowno{2}] {mozart-cl-s11.11.dat};
      \addlegendentry{$d_{2,\delta}^{q,\mathrm{rad}}(Y,y^{\mathrm{dir}})$}
      \addplot[thick,densely dashed,red] table
              [x expr=\thisrowno{0},
                y expr=\thisrowno{1}] {mozart-cl-nob-s11.11.dat};
      \addlegendentry{$d_{2,\delta}^{q,\mathrm{abs}}(Y,y^{\mathrm{spr}})$ ($b=0$)}
      \addplot[thick,densely dotted,red] table
              [x expr=\thisrowno{0},
                y expr=\thisrowno{2}] {mozart-cl-nob-s11.11.dat};
      \addlegendentry{$d_{2,\delta}^{q,\mathrm{rad}}(Y,y^{\mathrm{dir}})$ ($b=0$)}
    \end{axis}
    \end{tikzpicture}
    \caption{Observed loss convergence on the isolated piano track}
    \label{fig:piano}
\end{figure}

\subsection{URMP}

The URMP dataset \cite{Li18} consists of samples with two or more
acoustic instruments.  It was not created for blind separation, so
the samples are generally too \enquote{hard} to be used in that
context.  Nevertheless, in \cite{Schulze21}, a subset of potentially
appropriate samples was determined, and we compare the performance
of our new algorithm on these samples.

\begin{specialtable}[ht]
  \centering
  \caption{Comparison of the separation algorithms on a
    selection of samples from the URMP \cite{Li18} dataset.
    Best numbers are marked.}
  \label{tab:urmp}
  \tabcolsep=4.5pt
  \renewcommand{\arraystretch}{1.1}
  \heavyrulewidth=.5pt
  \lightrulewidth=.4pt
  \cmidrulewidth=.3pt
  \aboverulesep=0.0ex
  \belowrulesep=0.0ex

  \newcolumntype Y{S[
      table-format=2.1,
      table-auto-round,
      table-text-alignment=center,
      table-number-alignment=center,
      table-space-text-post={*}]}
  \newcolumntype Z{S[
      table-format=-1.1,
      table-auto-round,
      table-text-alignment=center,
      table-number-alignment=center,
      table-space-text-post={*}]}
    \begin{tabular}{ccZYZ}
      \toprule
    {\bfseries Method} & {\bfseries Instrument} & {\bfseries SDR} & {\bfseries SIR} & {\bfseries SAR} \\\midrule
    \multirow{8}{*}{Ours} & Flute & -4.7 & 17.5 & -4.6 \\
    & Clarinet & 5.0 & 10.1 & 7.0* \\
    \cmidrule{2-5}
    & Trumpet & 7.7* & 19.9* & 8.0* \\
    & Violin & 9.7* & 30.7* & 9.7* \\
    \cmidrule{2-5}
    & Trumpet & 8.4* & 30.3* & 8.4* \\
    & Saxophone & 13.0* & 24.9* & 13.3* \\
    \cmidrule{2-5}
    & Oboe & 2.9 & 6.9 & 5.9 \\
    & Cello & -0.6 & 19.2* & -0.5 \\
    \cmidrule{1-5}
    \multirow{8}{*}{\cite{Schulze21}} & Flute & 2.44635917 & 9.46831118 & 3.87247707* \\
    & Clarinet & 6.22754594* & 25.27585464* & 6.29482429 \\
    \cmidrule{2-5}
    & Trumpet & 5.27036917 & 16.55001473 & 5.70157274 \\
    & Violin & 7.70812429 & 25.11747363 & 7.80105425 \\
    \cmidrule{2-5}
    & Trumpet & -2.40914079 & 1.07264129 & 2.68289795 \\
    & Saxophone & 0.12786161 & 22.48902291 & 0.17756585 \\
    \cmidrule{2-5}
    & Oboe & 6.31074976* & 17.04997418* & 6.7782757* \\
    & Cello & 4.19790633* & 17.10544068 & 4.50990864 \\
    \cmidrule{1-5}
    \multirow{8}{*}{\cite{Duan08}} & Flute & 3.4085* & 19.5506* & 3.5633 \\
    & Clarinet & 2.0911 & 5.9174 & 5.4046 \\
    \cmidrule{2-5}
    & Trumpet & {---} & {---} & {---} \\
    & Violin & {---} & {---} & {---} \\
    \cmidrule{2-5}
    & Trumpet & 1.1638 & 9.4487 & 2.3286 \\
    & Saxophone & 6.9220 & 17.2436 & 7.4265 \\
    \cmidrule{2-5}
    & Oboe & -0.7868 & 13.1102 & -0.3989 \\
    & Cello & 3.3892 & 6.4313 & 7.2580* \\
    \bottomrule
  \end{tabular}
\end{specialtable}

As can be seen in Table \ref{tab:urmp}, the results vary widely.
For the first and the fourth sample, we have to declare a failure
compared to the other two algorithms.  On the second and the third
sample, however, our algorithm is universally dominant.
The good performance on the sample with trumpet and saxophone is
especially surprising, since we deemed this a very challenging
sample due to the similarity of the sounds of the instruments.

\subsubsection{Oracle Dictionary}

In order to investigate the failure of the separation method on the
first and the fourth sample, we first train \emph{oracle} dictionaries
by providing the algorithm with the ground-truth individual tracks for
the respective instruments.
We then supply the separation procedures with these dictionaries
as initial values.  In one instance, we keep the dictionaries fixed
throughout the training, and in another one, we train then at the normal rate,
starting from oracle dictionaries.
For each separation, we use $4$ different random seeds.  The results are
displayed in Table~\ref{tab:urmporacle}.  While we usually use the direct
prediction $z^{\mathrm{dir}}[k,l]$ for resynthesis, we here also include the
resynthesis based on the dictionary prediction $z[k,l]$ for analysis.

\begin{specialtable}[ht]
  \centering
  \caption{Separation with an oracle dictionary on a
    selection of samples from the URMP \cite{Li18} dataset.
    The \enquote{Fix} column indicates whether the dictionary is kept
    constant during the separation, and the \enquote{Pred.} column
    specifies whether the direct or the dictionary prediction
    is used.
    Best numbers are marked when they also exceed the performance
    from Table \ref{tab:urmp}.}
  \label{tab:urmporacle}
  \tabcolsep=4.5pt
  \renewcommand{\arraystretch}{1.1}
  \heavyrulewidth=.5pt
  \lightrulewidth=.4pt
  \cmidrulewidth=.3pt
  \aboverulesep=0.0ex
  \belowrulesep=0.0ex

  \newcolumntype Y{S[
      table-format=2.1,
      table-auto-round,
      table-text-alignment=center,
      table-number-alignment=center,
      table-space-text-post={*}]}
  \newcolumntype Z{S[
      table-format=-1.1,
      table-auto-round,
      table-text-alignment=center,
      table-number-alignment=center,
      table-space-text-post={*}]}
    \begin{tabular}{cccZYZ}
      \toprule
    {\bfseries Fix} & {\bfseries Pred.} & {\bfseries Instrument} & {\bfseries SDR} & {\bfseries SIR} & {\bfseries SAR} \\\midrule
    \multirow{8}{*}{Yes} & \multirow{4}{*}{Dir.} & Flute & 1.2 & 9.4 & 2.4 \\
    & & Clarinet & 5.7 & 25.7 & 5.8 \\
    \cmidrule{3-6}
    & & Oboe & 5.3 & 11.2 & 6.8* \\
    & & Cello & 3.0 & 30.3 & 3.0 \\
    \cmidrule{2-6}
    & \multirow{4}{*}{Dict.} & Flute & -0.5 & 1.01 & 0.3 \\
    & & Clarinet & 1.8 & 30.2* & 1.8 \\
    \cmidrule{3-6}
    & & Oboe & 0.5 & 9.6 & 1.6 \\
    & & Cello & -1.4 & 25.4 & -1.4 \\
    \cmidrule{1-6}
    \multirow{8}{*}{No} & \multirow{4}{*}{Dir.} & Flute & -0.4 & 21.6 & -0.3 \\
    & & Clarinet & 7.0* & 13.2 & 8.4* \\
    \cmidrule{3-6}
    & & Oboe & 3.7 & 8.0 & 6.3 \\
    & & Cello & 0.4 & 26.1* & 0.5 \\
    \cmidrule{2-6}
    & \multirow{4}{*}{Dict.} & Flute & -5.1 & 24.6* & -5.1 \\
    & & Clarinet & 2.2 & 17.3 & 2.4 \\
    \cmidrule{3-6}
    & & Oboe & -1.8 & 4.6 & 0.6 \\
    & & Cello & -2.8 & 23.7 & -2.8 \\
    \bottomrule
  \end{tabular}
\end{specialtable}

With the fixed oracle dictionary, the results are generally much better than
with the normal training in Table \ref{tab:urmp}.  However, when allowing
training from the oracle dictionary,
the separated flute and the cello tracks become unacceptably bad again.
When using the dictionary prediction for resynthesis, all the results
are of very poor quality, indicating that the dictionary model
\eqref{eq:dictmodel} is not an accurate representation of the spectral
characteristics of the tones.

Upon manual inspection of the flute track, we noticed that it contains a
number of tones which are \enquote{half-overblown,} such that the spectra
of both the higher octave and of the lower octave are present.  This does
not represent the normal spectral characterics of the flute sound, so
the oracle dictionary contains a \enquote{compromise,} while the trained
dictionary fails to represent these half-overblown tones.
In the cello track, there is no obvious technical peculiarity, but the
tones are simply very diverse, involving different open strings and
also different articulation between the tones, so even the training
of the oracle dictionary is problematic.

Generally, for instruments with inconsistent spectral characterics,
the method from \cite{Schulze21} may be at an advantage since it
prunes and randomly reinitializes parts of the dictionary in regular
intervals; so, given enough tries, it can reach an appropriate dictionary
by chance, even if it is potentially suboptimal with respect to the loss
function.

\subsection{Duan et al.}

In \cite{Duan08}, a number of original samples are used.  As we mentioned,
their algorithm has the unique ability of separating a representable
instrument track out of a mixture with a non-representable residual,
which can, for instance, be a singing voice.
Since our algorithm is not designed for such signals, we selected
the samples for which all the instruments can be represented.
In total, these are one sample with acoustic oboe and euphonium,
one sample with \emph{synthetic} piccolo and organ, and a third sample
with a synthetic oboe track added to the previous sample.

The problems with these samples are that they have a different sampling
frequency ($f_{\mathrm{s}}=\SI{22.05}{kHz}$) and they are also very short.
Whereas in \cite{Schulze21}, the signals were converted to a different
sampling frequency as a preprocessing step in order to reduce the loss of
resolution due to smoothing, we do not have this problem here.
While we keep the value of $\zeta$ from \eqref{eq:default}
constant in terms of absolute units, the relation to the sampling
frequency consequently changes to $\zeta f_{\mathrm{s}}=470.4$.  The frequency
constant changes proportionately with the sampling frequency to
$\beta=\SI{1.79443359375}{Hz}$.
Since the samples are short, we choose an even smaller time constant
compared to \eqref{eq:default}
by setting $\alpha=16/f_{\mathrm{s}}\approx\SI{0.73}{ms}$
in the acoustic sample and
$\alpha=128/f_{\mathrm{s}}\approx\SI{5.80}{ms}$
in the synthetic ones, each with
$\tilde{\alpha}=\alpha/4$ for training.  The results are displayed in
Table \ref{tab:duan}.

\begin{specialtable}[t]
  \centering
  \caption{Comparison of the separation algorithms on
    the data by \cite{Duan08}.  Instruments
    labeled as \enquote{s.} are synthetic, those labeled as
    \enquote{a.} are acoustic.  Best numbers are marked.}
  \label{tab:duan}
  \tabcolsep=4.5pt
  \renewcommand{\arraystretch}{1.1}
  \heavyrulewidth=.5pt
  \lightrulewidth=.4pt
  \cmidrulewidth=.3pt
  \aboverulesep=0.0ex
  \belowrulesep=0.0ex

  \newcolumntype Y{S[
      table-format=2.1,
      table-auto-round,
      table-text-alignment=center,
      table-number-alignment=center,
      table-space-text-post={*}]}
    \begin{tabular}{ccYYY}
      \toprule
    {\bfseries Method} & {\bfseries Instrument} & {\bfseries SDR} & {\bfseries SIR} & {\bfseries SAR} \\\midrule
    \multirow{7}{*}{Ours} & Oboe (a.) & 9.6 & 47.2* & 9.6 \\
    & Euphonium (a.) & 8.7 & 33.7* & 8.7 \\
    \cmidrule{2-5}
    & Piccolo (s.) & 17.2* & 36.5* & 17.2* \\
    & Organ (s.) & 14.3* & 50.3* & 14.3* \\
    \cmidrule{2-5}
    & Piccolo (s.) & 6.8* & 22.1 & 6.9* \\
    & Organ (s.) & 7.3* & 19.2 & 7.7* \\
    & Oboe (s.) & 8.3 & 46.3* & 8.3 \\
    \cmidrule{1-5}
    \multirow{7}{*}{\cite{Schulze21}} & Oboe (a.) & 18.63132098* & 33.5532616 & 18.77536405* \\
    & Euphonium (a.) & 14.65085084* & 31.46362964 & 14.74537616* \\
    \cmidrule{2-5}
    & Piccolo (s.) & 11.17925459 & 25.93718057 & 11.33799989 \\
    & Organ (s.) & 10.10624241 & 20.73031387 & 10.5362565 \\
    \cmidrule{2-5}
    & Piccolo (s.) & 4.22638963 & 24.79923549* & 4.27897973 \\
    & Organ (s.) & 6.04851133 & 19.96833418* & 6.27182761 \\
    & Oboe (s.) & 5.25789918 & 12.43157896 & 6.42346892 \\
    \cmidrule{1-5}
    \multirow{7}{*}{\cite{Duan08}} & Oboe (a.) & 8.7 & 25.8 & 8.8 \\
    & Euphonium (a.) & 4.6 & 14.5 & 5.3 \\
    \cmidrule{2-5}
    & Piccolo (s.) & 14.2 & 27.9 & 14.4 \\
    & Organ (s.) & 11.8 & 25.1 & 12.1 \\
    \cmidrule{2-5}
    & Piccolo (s.) & 6.5 & 20.0 & 6.7 \\
    & Organ (s.) & 6.6 & 17.3 & 7.1 \\
    & Oboe (s.) & 9.0* & 21.9 & 9.2* \\
    \bottomrule
  \end{tabular}
\end{specialtable}

While the results for the acoustic sample are better than in the original
publication, they are still not nearly as good as those in \cite{Schulze21}.
Our explanation is that while the time resolution of the spectrogram
is almost as high as that of the time-domain signal
($\tilde{\alpha}=4/f_{\mathrm{s}}$), there is still just not enough data
in the sample to train the neural network, and thus hand-crafted methods
are at an advantage.

By contrast, our method delivers very good results with synthetic
instruments, clearly and universally outperforming
the other methods on the sample with two instruments and providing
the best average performance on the sample with three instruments.

\section{Conclusion}

We have developed a blind source separation method that unmixes the
contributions of different instruments in a polyphonic music recording
via a parametric model and a dictionary.  The model parameters are predicted
by a deep convolutional neural network, and with respect to those that
do not possess a useful backpropagation gradient, we use the policy gradient
instead.

Unlike other algorithms, ours operates directly on the complex output of the
STFT, which is linear and preserves the phase.
Rather than using spectral masking, we
let the network give a direct prediction for the complex amplitudes of
the harmonics.

In general, the algorithm exhibits very good performance on a variety
of samples.  It is especially dominant in terms of SIR, which is relevant
since eliminating cross-talk is the main objective in separation.
We attribute this to
the use of a complex-valued direct prediction for the individual instruments
that can properly handle interference between the instrument tones in the
spectral domain.  Such interference is particularly prevalent in synthetic
samples, on which the performance of our algorithm surpasses that of
the competing methods.  It also clearly outperforms the other methods on
the sample with the acoustic piano, which poses the challenge of
detecting tones with inharmonicity.

As is usual with blind separation, however, problems arise when the
structural assumptions are not satisfied.  In two samples from the URMP
database, there was one instrument each whose sound in the recording
varied too much to be accurately represented by the dictionary.  While
we found that using oracle dictionaries, satisfactory separation can
be achieved, these dictionaries are not attained as the result of training,
even when they are supplied as the initial value.

Due to the use of neural networks, our approach is very flexible with respect
to the choice of the loss function.  In the spirit of blind separation,
we chose the weights for the respective distances such that they constitute
a reasonable comprise for all the samples on which we tested them,
but this choice is not necessarily optimal for the individual samples.
While a linear combination is the most straight-forward way to account
for all the considered distances, a non-linear mapping could potentially
be better.  Also, even though we have not found any distance functions
yielding better performance than the ones we use, more experiments
could be conducted.

With our approach, we hope to provide a blueprint for the combined
use of backpropagation gradients and policy gradients in the application of
neural networks on non-convex parameter identification problems.

\vspace{6pt} 

\authorcontributions{S.S. devised, implemented, and tested the algorithm.
S.S. and J.L. wrote the manuscript and conceived additional experiments.
E.J.K. supervised the research and revised the manuscript.
All authors have read and agreed to the published version of the manuscript}

\funding{S.S. and J.L. acknowledge funding by the Deutsche Forschungsgemeinschaft (DFG, German Research Foundation) -- project number 281474342/GRK2224/1.}

\institutionalreview{Not applicable}

\informedconsent{Not applicable}

\dataavailability{The functioning source code for the algorithm is available on
  GitHub\footnote{\url{https://github.com/rgcda/Musisep}} under the
  GNU General Public License (version 3).  The input data from \cite{Duan08}
  can be downloaded from the respective website\footnote{%
  \url{https://sites.google.com/site/mperesult/musicseparationresults}}.
  For all the other samples, we provide the input data along with
  the best-case separation results on the institute website\footnote{%
  \url{https://www.math.colostate.edu/~king/software.html\#Musisep}}.
  Due to the stochastic nature of parallel computing, training results are not
  exactly reproducible.}

\acknowledgments{The authors would like to thank Sören Dittmer and
Louisa Kinzel for sharing their expertise and ideas on the learning approach.}

\conflictsofinterest{The authors declare no conflict of interest.} 

\abbreviations{Abbreviations}{
The following abbreviations are used in this manuscript:\\

\noindent
\begin{tabular}{@{}lp{10cm}}
  DIP & Deep image prior\\
  GAN & Generative adversarial network\\
  MCTS & Monte Carlo tree search\\
  NMF & Non-negative matrix factorization\\
  PLCA & Probabilistic latent component analysis\\
  REINFORCE & $\text{Reward increment} = \text{Nonnegative factor} \times \text{Offset reinforcement}$\newline $\null\times \text{Characteristic eligibility}$\\
  SAR & Signal-to-artifacts ratio\\
  SDR & Signal-to-distortion ratio\\
  SIR & Signal-to-interference ratio\\
  STFT & Short-time Fourier transform\\
  URMP & University of Rochester Multi-Modal Music Performance Dataset
\end{tabular}}

\end{paracol}
\reftitle{References}

\externalbibliography{yes}
\bibliography{paper.bib}

\end{document}